\documentclass[12pt]{article}
\usepackage{amsmath}
\usepackage{graphicx,psfrag,epsf}
\usepackage{enumerate}
\usepackage{natbib}
\usepackage{textcomp}
\usepackage[hyphens]{url} 
\usepackage{hyperref}

\newcommand{\blind}{0}

\usepackage{color}
\usepackage{fancyvrb}

\DefineVerbatimEnvironment{Highlighting}{Verbatim}{commandchars=\\\{\}}
\usepackage{framed}
\definecolor{shadecolor}{RGB}{248,248,248}
\newenvironment{Shaded}{\begin{snugshade}}{\end{snugshade}}

\newcommand{\AttributeTok}[1]{\textcolor[rgb]{0.77,0.63,0.00}{#1}}

\newcommand{\CommentTok}[1]{\textcolor[rgb]{0.56,0.35,0.01}{\textit{#1}}}

\newcommand{\ConstantTok}[1]{\textcolor[rgb]{0.00,0.00,0.00}{#1}}
\newcommand{\ControlFlowTok}[1]{\textcolor[rgb]{0.13,0.29,0.53}{\textbf{#1}}}

\newcommand{\DecValTok}[1]{\textcolor[rgb]{0.00,0.00,0.81}{#1}}

\newcommand{\FloatTok}[1]{\textcolor[rgb]{0.00,0.00,0.81}{#1}}
\newcommand{\FunctionTok}[1]{\textcolor[rgb]{0.00,0.00,0.00}{#1}}

\newcommand{\NormalTok}[1]{#1}

\newcommand{\OtherTok}[1]{\textcolor[rgb]{0.56,0.35,0.01}{#1}}

\newcommand{\SpecialCharTok}[1]{\textcolor[rgb]{0.00,0.00,0.00}{#1}}

\newcommand{\StringTok}[1]{\textcolor[rgb]{0.31,0.60,0.02}{#1}}

\providecommand{\tightlist}{%
  \setlength{\itemsep}{0pt}\setlength{\parskip}{0pt}}

\newcommand{\github}[1]{\href{https://github.com/mine-cetinkaya-rundel/educators-perspective-tidyverse}{https://github.com/mine-cetinkaya-rundel/educators-perspective-tidyverse}}

\usepackage{xcolor}
\xdefinecolor{blue}{rgb}{0.28,0.65,0.9}
\hypersetup{
    colorlinks,
    linkcolor={blue},
    citecolor={blue},
    urlcolor={blue}
}

\usepackage[margin=0.6cm]{caption}
\usepackage{floatrow}

\DeclareNewFloatType{chunk}{placement=H, fileext=chk, name=}
\makeatletter
\@addtoreset{chunk}{section}
\makeatother
\usepackage{xcolor,soul}
\usepackage{xspace}
\usepackage{booktabs}
\usepackage{longtable}
\usepackage{array}
\usepackage{multirow}
\usepackage{wrapfig}
\usepackage{float}
\usepackage{colortbl}
\usepackage{pdflscape}
\usepackage{tabu}
\usepackage{threeparttable}
\usepackage{threeparttablex}
\usepackage[normalem]{ulem}
\usepackage{makecell}
\usepackage{xcolor}

\begin{document}

\def\spacingset#1{\renewcommand{\baselinestretch}%
{#1}\small\normalsize} \spacingset{1}


\if0\blind
{
  \title{\bf An educator's perspective of the tidyverse}

  \author{
        Mine \c{C}etinkaya-Rundel \footnote{\href{mailto:mc301@duke.edu}{\nolinkurl{mc301@duke.edu}}} \\
    Department of Statistical Science, Duke University and RStudio\\
     and \\     Johanna Hardin \footnote{\href{mailto:jo.hardin@pomona.edu}{\nolinkurl{jo.hardin@pomona.edu}}} \\
    Department of Mathematics and Statistics, Pomona College\\
     and \\     Benjamin S. Baumer \footnote{\href{mailto:bbaumer@smith.edu}{\nolinkurl{bbaumer@smith.edu}}} \\
    Program in Statistical \& Data Sciences, Smith College\\
     and \\     Amelia McNamara \footnote{\href{mailto:amelia.mcnamara@stthomas.edu}{\nolinkurl{amelia.mcnamara@stthomas.edu}}} \\
    Department of Computer \& Information Sciences, University of St
Thomas\\
     and \\     Nicholas J. Horton \footnote{\href{mailto:nhorton@amherst.edu}{\nolinkurl{nhorton@amherst.edu}}} \\
    Department of Mathematics and Statistics, Amherst College\\
     and \\     Colin W. Rundel \footnote{\href{mailto:colin.rundel@duke.edu}{\nolinkurl{colin.rundel@duke.edu}}} \\
    Department of Statistical Science, Duke University\\
      }
  \maketitle
} \fi

\if1\blind
{
  \bigskip
  \bigskip
  \bigskip
  \begin{center}
    {\LARGE\bf An educator's perspective of the tidyverse}
  \end{center}
  \medskip
} \fi

\bigskip
\begin{abstract}
Computing makes up a large and growing component of data science and
statistics courses. Many of those courses, especially when taught by
faculty who are statisticians by training, teach R as the programming
language. A number of instructors have opted to build much of their
teaching around use of the \textbf{tidyverse}. The tidyverse, in the
words of its developers, ``is a collection of R packages that share a
high-level design philosophy and low-level grammar and data structures,
so that learning one package makes it easier to learn the next''
\citep{tidyverse}. These shared principles have led to the widespread
adoption of the tidyverse ecosystem. A large part of this usage is
because the tidyverse tools have been intentionally designed to ease the
learning process and make it easier for users to learn new functions as
they engage with additional pieces of the larger ecosystem. Moreover,
the functionality offered by the packages within the tidyverse spans the
entire data science cycle, which includes data import, visualisation,
wrangling, modeling, and communication. We believe the tidyverse
provides an effective and efficient pathway for undergraduate students
at all levels and majors to gain computational skills and thinking
needed throughout the data science cycle. In this paper, we introduce
the tidyverse from an educator's perspective. We provide a brief
introduction to the tidyverse, demonstrate how foundational statistics
and data science tasks are accomplished with the tidyverse, and discuss
the strengths of the tidyverse, particularly in the context of teaching
and learning.
\end{abstract}

\noindent%
\vfill

\newpage
\spacingset{1.45} 

\hypertarget{introduction}{%
\section{Introduction}\label{introduction}}

Computing has a fundamental and growing role in the statistics and data
science curriculum \citep{horthard_2021, nolan_templelang_2010}. The
revised Guidelines for Assessment and Instruction in Statistics
Education (GAISE) College report notes the importance of technology and
states: ``ideally, students should be given numerous opportunities to
analyze data with the best available technology (preferably, statistical
software)'' (\citet{CarEve2016}, page 11). In both statistics and data
science courses, we believe it is important to teach tools that are used
by practitioners of these disciplines (i.e., authentic tools).

When it comes to an authentic tool, \citet{mcna_2019} argues that a
modern statistical computing tool ``should be accessible, provide easy
entry, privilege data as a first-order object, support exploratory and
confirmatory analysis, allow for flexible plot creation, support
randomization, be interactive, include inherent documentation, support
narrative, publishing, and reproducibility, and be flexible to
extensions'' (page 1). Such tools ideally allow new users and
professionals to ``reach across the gap'' between tools for teaching and
tools for doing to foster continued learning \citep{mcnamara2015}. R
exhibits all of these attributes, particularly with careful curation and
thoughts toward pedagogy. It is also flexible, powerful, and
open-source. As a result, many instructors at the undergraduate level,
particularly those with a background in statistics, have chosen R
\citep{Rlang} for their teaching.

There are many pedagogical decisions that emerge when an instructor
chooses to teach with a particular computational platform or tool. We
describe how the \textbf{tidyverse}, a collection of packages intended
to provide a consistent interface in R, reduces friction for both the
instructor and the student across the entire data analysis cycle, which
is foundational to both statistics and data science. Like many other
instructors, we have opted to build much of our teaching around use of
the tidyverse. This paper is a synthesis of the reasoning for our
choice, along with benefits and challenges associated with teaching
(with) the tidyverse.

We begin with a description of the tidyverse (``what'') in Section
\ref{sec:principles}, including the design principles that guide its
development and promote ease of learning. Section \ref{sec:tidyverse}
follows with examples of ``how'' the tidyverse works, including
comparisons of the tidyverse approach with base R approaches and an
outline of core packages and functions. Section \ref{sec:teach}
articulates ``why'' one might teach using this approach: namely because
it is consistent, scalable, user-centered, readable, and popular.
Section \ref{sec:discussion} provides closing thoughts and discussion.

This paper focuses primarily on undergraduate introductory statistics
and data science courses. However, we will also comment on how gaining
an introduction to R and the tidyverse in these courses may help prepare
a student for success in higher level courses at the undergraduate and
graduate levels, as well as in industry.

There are several popular, free, open-source, programming languages that
can be used in introductory statistics and data science, including R
\citep{Rlang}, Python \citep{CS-R9526}, and Julia
\citep{mcnicholas2019data}. These languages also display many attributes
promulgated by \citet{mcna_2019}. We note that the influential
\href{http://data8.org/}{Data 8} course at the University of California,
Berkeley (as well as the follow-up \href{https://ds100.org/sp21/}{Data
100} course) are taught in Python, with a significant portion of the
instruction centered around a course-specific Python library. We affirm
that students benefit from developing literacy in multiple languages and
argue that the ``tidy data'' \citep{wickham2014tidy} approach central to
the tidyverse is programming language independent, with notable
implementations in the three languages mentioned above as well as domain
specific languages like SQL (see Section \ref{sec:shared-syntax}). We
have chosen to focus our attention on R, in part because there are good
models for teaching statistics and data science with reproducible
computing practices---even at the introductory level
\citep{baumer2014, beckman2021implementing}.

Instructors teaching R face a pedagogical decision about how to teach
it. Some instructors use the ``I'll just teach it how I learned it''
approach, which we assert is not sound pedagogical reasoning. Chances
are that many things have changed since the time you first learned R.
All the authors have seen major changes to R over their careers, even
the most junior among us. As with the need to keep up with all
curricular and pedagogical changes in statistics \citep{zieffler2008},
\citet{nolan_templelang_2010} highlight the importance of workshops or
other modes for providing instructors with the skills to teach modern
statistical computing. One of the exciting, albeit challenging, aspects
of teaching R (or any computing language or tool) is that the landscape
is continuously evolving. While a changing landscape means instructors
need to continue learning, it also means that R has become more
user-friendly and student-friendly over time. Recent developments,
perhaps most notably the tidyverse, have helped to round off rough edges
and make R interfaces and syntax more coherent and consistent. It is
important for educators to periodically reevaluate their teaching in
light of what is most widely used, what is more user-friendly, what has
better documentation, what has better learning resources, what has
better community support, etc.

Based on all of these considerations (and more that we will articulate
below), and despite the need to incorporate additional learning outcomes
into our classes, we recommend teaching with the tidyverse as a way to
further integrate computation into our courses and programs. While we
recommend the use of the RStudio integrated development environment
(IDE; \citet{rstudio}), our arguments for the tidyverse also stand
outside the RStudio IDE. Others have focused on how using R to teach
statistics can decrease the cognitive load (amount of information needed
at once) of the class
\citep{Pruim2017, guzman2019successful, tuckerstatistics2021}, which we
affirm from our personal teaching experiences. Here, we focus on the
many different pedagogical advantages of using the tidyverse and the
specific benefits they bring to the statistics and data science
classroom. Table \ref{tab:summary} summarizes our arguments.

\linespread{1}
\begin{table}

\caption{\label{tab:tab:summary}Summary of pedagogical benefits of the tidyverse discussed in this paper.\label{tab:summary}}
\centering
\begin{tabular}[t]{l|>{\raggedright\arraybackslash}p{28em}}
\hline
Concept & Description\\
\hline
Consistency & Syntax, function interfaces, argument names and orders follow patterns\\
\hline
Mixability & Ability to use base and other functions within tidyverse syntax\\
\hline
Scalability & Unified approach to data wrangling and visualization works for datasets of a wide range of types and sizes\\
\hline
User-centered design & Function interfaces designed with users in mind\\
\hline
Readability & Interfaces that are designed to produce readable code\\
\hline
Community & Large, active, welcoming community of users and resources\\
\hline
Transferability & Data manipulation verbs inherit from SQL's query syntax\\
\hline
\end{tabular}
\end{table}\linespread{2}
\vspace{3mm}\setlength{\parindent}{15pt}

\hypertarget{sec:principles}{%
\section{Principles of the tidyverse}\label{sec:principles}}

\begin{figure}

{\centering \includegraphics[width=0.8\linewidth]{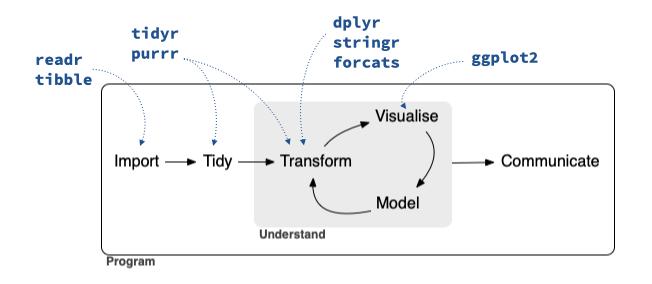} 

}

\caption{Data science cycle and the core tidyverse packages that address each phase.}\label{fig:data-science-cycle}
\end{figure}

The tidyverse package is a meta R package that loads eight core packages
when invoked, and also bundles numerous other packages upon installation
(as suggested dependencies). These packages all share a design
philosophy as well as common grammar and data structures. The core
packages and the phases of the data science cycle they address are shown
in Figure \ref{fig:data-science-cycle}.

\linespread{1}
\linespread{2}
\vspace{3mm}\setlength{\parindent}{15pt}

\citet{tidyverse-style} outlines the design principles of the packages
in the tidyverse as the following:

\begin{itemize}
\tightlist
\item
  human-centered design: The tidyverse is designed specifically to
  support the activities of a human data analyst.
\item
  consistency: The functions in tidyverse packages are designed with a
  consistent interface, which allows the user to apply what they learned
  in one function to another and minimizes the number of special cases
  one needs to remember.
\item
  composability: The tidyverse functions allow users to solve complex
  problems by breaking them down into small pieces, many of which can be
  combined using the ``pipe'' operator (or a ``+'' operator for
  \textbf{ggplot2}). They support iterative exploratory analysis to find
  the best solution.
\item
  inclusivity: In addition to the packages and functions themselves, the
  tidyverse also refers to the community of people who use the packages.
  The packages, their documentation, and the community support all
  emphasize inclusivity.
\end{itemize}

Central to the tidyverse is the notion of tidy data, which
\citet{wickham2014tidy} defines as ``easy to manipulate, model and
visualize, and have a specific structure: each variable is a column,
each observation is a row, and each type of observational unit is a
table'' (page 1).

Many of the core packages in the tidyverse were originally developed
independently to address specific phases of the data science cycle, and
subsequently came together under the tidyverse umbrella in 2016
\citep{smithdavid2016}. All of the core tidyverse packages are under
active development, as the authors iterate to improve functionality.

Many, though not all, of the developers of the tidyverse packages are
funded by RStudio, PBC and work full-time on their open-source
development. The packages are co-developed with community contributors
and released under the open source MIT license. Thus, while development
of the tidyverse is funded by RStudio, the packages do not \emph{belong}
to RStudio.

\hypertarget{starting-with-a-tidy-data-frame}{%
\subsection{Starting with a (tidy) data
frame}\label{starting-with-a-tidy-data-frame}}

In R, data often live in a data frame whose columns represent variables
that we want to analyze. While the simple data frame structure is
fundamental to understanding data science and statistics, working with
it can be challenging for new learners.

When such a dataset is loaded into R, it is available as an object
called a data frame. When using base R the variables in that data frame
are commonly accessed with the \texttt{\$} operator (e.g.,
\texttt{loans\$loan\_amount} to access the variable called
\texttt{loan\_amount} in the data frame \texttt{loans}). Often, students
are tempted to access the \texttt{loan\_amount} variable in this example
by referring to it simply as \texttt{loan\_amount} and not specifying
the name of the data frame in which it lives. This results in a
frustrating error:
\texttt{object\ \textquotesingle{}loan\_amount\textquotesingle{}\ not\ found}.
Students experience this misconception when they think about variables
as stand-alone objects as opposed to components of the data frame in
which they live.

One approach to addressing the data frame versus variable challenge in
base R is using the \texttt{attach()} function, which makes the
variables in a data frame available in the global environment. Copying
variables into the global environment is not recommended practice
\citep{google-style} as it can cause name collisions when data frames
with identically named variables are attached in the same environment
(e.g., if you also had a \texttt{car\_loans} data frame that also
happened to have a variable called \texttt{loan\_amount} attached).
Additionally, using \texttt{attach()} to include a data frame's columns
in the global namespace can muddle understanding of connections between
the dataset and the variable within it. Pedagogically, we want students
to understand how observations and variables are linked to a structured
dataset.

A second approach to working with variables in base R is using the
\texttt{with()} function, which evaluates an expression within a
specified environment (e.g.,
\texttt{with(loans\_full\_schema,\ loan\_amount)} instead of
\texttt{loans\_full\_schema\$loan\_amount}). Using \texttt{with()}
avoids the name collision problem introduced by \texttt{attach()} but it
is more verbose and requires understanding how a new function works just
to access a variable in a data frame.

In the tidyverse, the first argument of each function, whether for data
wrangling, visualization, or any of the more complex tasks that can be
introduced later, is almost always a data frame. Tidyverse functions
allow access to variables in the data frame without having to re-specify
the name of the data frame (e.g., \texttt{arrange(loans,\ loan\_amount)}
will arrange the rows of the \texttt{loans} data frame based on the
value of \texttt{loan\_amount}). By forcing users to work with tidy
data, the tidyverse emphasizes the connection between data frames and
variables, helping to underscore fundamental data science and statistics
concepts while also simplifying the syntax for routine data analysis
tasks.

The tidyverse strongly encourages the use of the pipe operator to
construct readable, vertical data pipelines, whether it's for data
wrangling or visualization. Beginning with R 4.1.0, the previous
\texttt{arrange()} code would be written as
\texttt{loans\ \textbar{}\textgreater{}\ arrange(loan\_amount)} using
the pipe operator (\texttt{\textbar{}\textgreater{}}) to pipe the data
frame (or any result from the preceding step of the pipeline) into the
subsequent function as its first argument (see Section \ref{sec:wrangle}
for more details on using the pipe operator in the
tidyverse).\footnote{In versions of R earlier than 4.1.0, the magrittr
  pipe (\texttt{\%\textgreater{}\%}) must be used in place of the native
  pipe (\texttt{\textbar{}\textgreater{}}). When using the base R pipe,
  note that \emph{all} students must be running R version 4.1 or newer).}
Pipe operators exist as a common feature across a number of other
programming languages (e.g., the UNIX shell, JavaScript, etc.) and were
introduced into the R ecosystem by the \textbf{magrittr} package
\citep{R-magrittr} package. The widespread community adoption of the
tidyverse (and the corresponding improved error recovery) led to the
introduction of a pipe operator into the base language\footnote{See, for
  example,
  \href{https://www.youtube.com/watch?v=X_eDHNVceCU\&t=4085s}{Luke
  Tierney's talk at the 2020 R Core useR! event}, in which he states
  that the R Core Team is considering adding a pipe operator to base R
  because because ``magrittr is very popular'' and ``a number of other
  languages are adding pipes.''} in early 2021 with R version 4.1.0.

Like the \texttt{lm()} function, or systems such as SAS (which feature a
common interface with a \texttt{DATA\ =} statement), tidyverse functions
take a \texttt{data} argument that allows them to localize computations
inside the specified data frame. The tidyverse approach is attractive
because it does not muddy the concept of what is in the current
environment (always the data as a data frame, never a variable as a
vector) while making it easy for variables in a data frame to be
accessed without the use of an additional function (like
\texttt{with()}) or even quotation marks. Furthermore, unlike
\texttt{lm()}, functions in the tidyverse almost always take a data
frame as their \emph{first} argument and \emph{return} a data frame. The
consistent structure of data frames and variables makes it easier to get
started with data analysis tasks without getting bogged down by language
details or using more complicated programming practices.

We note that data frames within the tidyverse are stored as
\texttt{tibble}s which have class \texttt{tbl\_df} in addition to
\texttt{data.frame}. \texttt{tibble}s act similarly to all other
\texttt{data.frame}s with less transformation of input (e.g., character
vectors are not coerced to factors, column names are not modified).
Additionally, when printed, by default only the first ten rows and as
many columns as fit into the current console or document are shown, and
the rest of the rows and columns are summarized. Conversely, when
\texttt{data.frame}s are printed all rows and columns are shown
regardless of how large (width or length-wise) the data are.

\hypertarget{consistent-grammar-and-vocabulary}{%
\subsection{Consistent grammar and
vocabulary}\label{consistent-grammar-and-vocabulary}}

\citet{hermans2017writing} used the metaphor of programming as writing,
and we extend their reasoning to assert that R is a programming language
with many syntaxes (or flavors/``dialects''). Different R packages can
use different syntaxes for the same idiom. Even packages included in the
distribution of base R do not all have a consistent grammar and
vocabulary. As stated in its design philosophy, the tidyverse strives
for consistency across packages \citep{tidyverse-style}. This makes the
tidyverse syntactically different from base R for doing certain tasks,
which might lead to learners of the tidyverse being less familiar with
base R code, and vice versa.

Tidyverse users tend to use particular vocabulary (e.g., pipes, tibbles,
verbs) compared to base R users who are more likely to speak in terms of
matrices, dollar signs (\texttt{\$}), and square brackets
(\texttt{{[}{]}}). A third commonly used approach is the formula syntax
\citep{Pruim2017}, which is characterized by the use of tildes
(\texttt{\textasciitilde{}}). The notion of a ``grammar'' in R code is
well-established for both graphics
\citep{wilkinson2012grammar, R-ggplot2} and data wrangling
\citep{R-dplyr}. In Section \ref{sec:reading}, we develop notions of
pronunciation. In spoken language, the word `dialect' refers to a
variety of language with distinct vocabulary, grammar, and
pronunciation, so we could consider the tidyverse as a dialect among
users who read and write R code.

No matter which approach or tool you use, you should strive to be
consistent in the classroom whenever possible. Our choice of the
tidyverse offers consistency, something we believe to be of the utmost
importance, allowing students to move knowledge about function arguments
to their long-term memory \citep{mcnamaraetal2021a}. From our
experience, switching between tools can lead to confusion for students,
and switching between syntaxes creates similar difficulties. Others have
highlighted the benefits of using a consistent and well-named syntax
\citep{Pruim2017, gehrke2021}.

One complication of teaching consistently is that Google and
StackOverflow can be less useful for students who are taught in only a
single modality. Searching online for answers is an important skill to
learn, but because of the variety of extant R syntaxes, searches can
lead students to patterns that are unfamiliar. Students should be
explicitly taught \emph{how} to search online and sift through results,
emphasizing the fact they are learning a specific syntax and only
responses in that syntax will make sense to them. Better searches will
include the names of specific tidyverse packages to each search query.
Being transparent and clear about the use of a syntax will help students
situate their knowledge---and misunderstandings---in the broader R
ecosystem.

One obvious counter-proposition is that you should teach \emph{all} (or
multiple) syntaxes at once. We strongly disagree. Trying to teach two
(or more!) syntaxes at once will slow the pace of the course, introduce
unnecessary syntactic confusion, and make it harder for students to
complete their work. In keeping with the ``let them eat cake first''
approach \citep{cetinkaya2020fresh, wang2017}, students benefit from
seeing the powerful things that they can do with R first. Deeper
discussions about syntax and under-the-hood programming concepts can
occur in subsequent courses after students are already invested in using
R. The approach we recommend is not tidyverse instead of base R, but
tidyverse (mostly) before base R.

We typically espouse a policy of being disciplined in what we teach,
liberal in what we accept \citep{postel1980dod}. In this paradigm,
instructors are careful about what they teach but choose an appropriate
level of flexibility in the code that students may submit. One might
adopt a policy that any code that works is acceptable. Another might
insist that only tidyverse patterns are acceptable. To continue the
analogy with language, a writing instructor might reasonably accept
papers written in any vernacular, or they might insist on a particular
writing style for a particular assignment.

\hypertarget{sec:tidyverse}{%
\section{Teaching foundational topics with the
tidyverse}\label{sec:tidyverse}}

In this section we provide comparative examples of data wrangling and
visualization tasks completed with the two most common R syntaxes: base
R and the tidyverse. The two tasks are essential to introductory data
science and introductory statistics courses as well as for data
practitioners. We believe that using the tidyverse makes these tasks
more straightforward than other approaches with R. Our approach is
similar to those presented in \citet{kleinman2009sas} for comparison of
SAS and R syntax; \citet{mcnamaraamelia2021} for comparison of base R,
tidyverse, and formula syntaxes in R; and \citet{dierker2021} for
comparisons across multiple statistical software programs.

\hypertarget{sec:wrangle}{%
\subsection{Data wrangling}\label{sec:wrangle}}

Data wrangling is a major component of data acumen \citep{nasem2018}
that often takes up a majority of the data analysis cycle. The tidyverse
includes a set of common idioms for data wrangling that work in a
consistent manner via the \textbf{dplyr} and \textbf{tidyr} packages.

\begin{itemize}
\tightlist
\item
  \texttt{filter()}: select rows
\item
  \texttt{select()}: select columns
\item
  \texttt{arrange()}: order rows
\item
  \texttt{mutate()}: add new or redefine existing columns
\item
  \texttt{group\_by()}: create partitions of rows
\item
  \texttt{summarize()}: aggregate (or ``roll up'') across rows
\item
  \texttt{*\_join()}: merge tables
\item
  \texttt{pivot\_*()}: reshape tables
\end{itemize}

While equivalents for each of these exist in base R and have been used
for decades, the base routines were developed independently over time
and do not share a common interface. In contrast, these functions always
take a data frame as a first argument, return a data frame, and use a
consistent naming convention for arguments.

To highlight some of our arguments, we will use the
\texttt{loans\_full\_schema} dataset from the
\href{http://openintrostat.github.io/openintro/index.html}{\textbf{openintro}}
package for code examples \citep{R-openintro}. The dataset represents
thousands of loans made through the Lending Club platform, which allows
individuals to lend to other individuals. The dataset contains
information on the applicants and their financial history. In
\ref{loans-data}, a small amount of data wrangling has been done to set
the dataset up for analyses throughout the paper. A fully reproducible
version of this paper, including the R code for reproducing all
examples, can be found on GitHub at \github{}.

\linespread{1}

\begin{Shaded}
\begin{Highlighting}[]
\FunctionTok{library}\NormalTok{(tidyverse)}
\NormalTok{loans }\OtherTok{\textless{}{-}}\NormalTok{ openintro}\SpecialCharTok{::}\NormalTok{loans\_full\_schema }\SpecialCharTok{|\textgreater{}}
  \FunctionTok{mutate}\NormalTok{(}
    \AttributeTok{homeownership =} \FunctionTok{str\_to\_title}\NormalTok{(homeownership), }
    \AttributeTok{bankruptcy =} \FunctionTok{if\_else}\NormalTok{(public\_record\_bankrupt }\SpecialCharTok{\textgreater{}=} \DecValTok{1}\NormalTok{, }\StringTok{"Yes"}\NormalTok{, }\StringTok{"No"}\NormalTok{)}
\NormalTok{  ) }\SpecialCharTok{|\textgreater{}}
  \FunctionTok{filter}\NormalTok{(annual\_income }\SpecialCharTok{\textgreater{}=} \DecValTok{10}\NormalTok{)}
\end{Highlighting}
\end{Shaded}

\captionof{chunk}{The \texttt{loans} dataset from the openintro package.}

\label{loans-data} \linespread{2}
\vspace{3mm}\setlength{\parindent}{15pt}

\ref{loans-data} provides a pipeline for: accessing a data frame within
the openintro package, creating two new variables using the
\texttt{mutate()} function, excluding incomes below \$10, and storing
the result in a new tibble called \texttt{loans}.

Suppose we want to compute the average income of applicants based on
their home ownership status. In the tidyverse, we could use the
following pipeline shown in \ref{tidy-summary}.

\linespread{1}

\begin{Shaded}
\begin{Highlighting}[]
\NormalTok{loans }\SpecialCharTok{|\textgreater{}}
  \FunctionTok{group\_by}\NormalTok{(homeownership) }\SpecialCharTok{|\textgreater{}}
  \FunctionTok{summarize}\NormalTok{(}
    \AttributeTok{num\_applicants =} \FunctionTok{n}\NormalTok{(),}
    \AttributeTok{avg\_loan\_amount =} \FunctionTok{mean}\NormalTok{(loan\_amount)}
\NormalTok{  ) }\SpecialCharTok{|\textgreater{}}
  \FunctionTok{arrange}\NormalTok{(}\FunctionTok{desc}\NormalTok{(avg\_loan\_amount))}
\end{Highlighting}
\end{Shaded}

\begin{verbatim}
## # A tibble: 3 x 3
##   homeownership num_applicants avg_loan_amount
##   <chr>                  <int>           <dbl>
## 1 Mortgage                4778          18132.
## 2 Own                     1350          15665.
## 3 Rent                    3848          14396.
\end{verbatim}

\captionof{chunk}{Using the tidyverse to count applicants and compute the average loan amount from the loans data, sorted by average loan amount.}

\label{tidy-summary} \linespread{2}
\vspace{3mm}\setlength{\parindent}{15pt}

The tidyverse syntax expresses the sequential process of the
computation. The pipe operator (\texttt{\textbar{}\textgreater{}})
brings the object from the left of the pipe into the function on the
right of the pipe as the first argument; we pronounce the pipe function
as ``and then.'' First, we start with the data frame that contains all
the data. And then, we group the data according to the unique values of
the \texttt{homeownership} variable. And then, for each unique value of
\texttt{homeownership}, we compute both the number of rows (calling the
result \texttt{num\_applicants}) and the average loan amount in US
Dollars (\texttt{avg\_loan\_amount}). And then, we arrange the rows of
the resulting data frame in \texttt{desc}ending order according to the
value of \texttt{avg\_loan\_amount}.

In base R, we might perform the same task with a computation similar to
the one in \ref{base-summary}.

\linespread{1}

\begin{Shaded}
\begin{Highlighting}[]
\NormalTok{res1 }\OtherTok{\textless{}{-}} \FunctionTok{aggregate}\NormalTok{(loan\_amount }\SpecialCharTok{\textasciitilde{}}\NormalTok{ homeownership, }\AttributeTok{data =}\NormalTok{ loans, }\AttributeTok{FUN =}\NormalTok{ length)}
\FunctionTok{names}\NormalTok{(res1)[}\DecValTok{2}\NormalTok{] }\OtherTok{\textless{}{-}} \StringTok{"num\_applicants"}
\NormalTok{res2 }\OtherTok{\textless{}{-}} \FunctionTok{aggregate}\NormalTok{(loan\_amount }\SpecialCharTok{\textasciitilde{}}\NormalTok{ homeownership, }\AttributeTok{data =}\NormalTok{ loans, }\AttributeTok{FUN =}\NormalTok{ mean)}
\FunctionTok{names}\NormalTok{(res2)[}\DecValTok{2}\NormalTok{] }\OtherTok{\textless{}{-}} \StringTok{"avg\_loan\_amount"}
\NormalTok{res }\OtherTok{\textless{}{-}} \FunctionTok{merge}\NormalTok{(res1, res2)}
\NormalTok{res[}\FunctionTok{order}\NormalTok{(res}\SpecialCharTok{$}\NormalTok{avg\_loan\_amount, }\AttributeTok{decreasing =} \ConstantTok{TRUE}\NormalTok{), ]}
\end{Highlighting}
\end{Shaded}

\begin{verbatim}
##   homeownership num_applicants avg_loan_amount
## 1      Mortgage           4778        18132.45
## 2           Own           1350        15665.44
## 3          Rent           3848        14396.44
\end{verbatim}

\captionof{chunk}{Using base R to count applicants and compute the average loan amount from the loans data, sorted by average loan amount.}

\label{base-summary} \linespread{2}
\vspace{3mm}\setlength{\parindent}{15pt}

We find the base R code harder to read and less expressive of the
logical process of the computation (i.e., more \emph{cryptic}). It
requires storing intermediate objects (\texttt{res1}, \texttt{res2}, and
\texttt{res}) that might not otherwise be useful. It uses the
\texttt{\textasciitilde{}}, \texttt{\$}, and \texttt{{[}} operators. It
uses a magic number (\texttt{2}) to hard-code the second variable. It
creates two data frames that need to be merged together. It passes the
name of a function as an argument to a function. In Section
\ref{sec:scalability}, we argue that the base R syntax for the task at
hand does not scale well for additional summary statistics. And in light
of Section \ref{sec:reading}, the base R syntax is quite challenging to
read aloud.

A different base R pattern, shown in \ref{tapply-summary}, is more
compact, and avoids some of the pitfalls listed above, but returns a
vector. This named vector makes the result easy to read. However, it
makes it cumbersome to include the second variable indicating the number
of people.

\linespread{1}

\begin{Shaded}
\begin{Highlighting}[]
\FunctionTok{sort}\NormalTok{(}\FunctionTok{tapply}\NormalTok{(loans}\SpecialCharTok{$}\NormalTok{loan\_amount, loans}\SpecialCharTok{$}\NormalTok{homeownership, mean), }\AttributeTok{decreasing =} \ConstantTok{TRUE}\NormalTok{)}
\end{Highlighting}
\end{Shaded}

\begin{verbatim}
## Mortgage      Own     Rent 
## 18132.45 15665.44 14396.44
\end{verbatim}

\captionof{chunk}{Using \texttt{tapply()} within base R to count applicants and compute the average loan amount from the loans data, sorted by average loan amount.}

\label{tapply-summary} \linespread{2}
\vspace{3mm}\setlength{\parindent}{15pt}

Piping can simplify the code, as seen in \ref{tapply-tidy-pipe}
(equivalent results with the native pipe not shown).

\linespread{1}

\begin{Shaded}
\begin{Highlighting}[]
\FunctionTok{tapply}\NormalTok{(loans}\SpecialCharTok{$}\NormalTok{loan\_amount, loans}\SpecialCharTok{$}\NormalTok{homeownership, mean) }\SpecialCharTok{|\textgreater{}}
  \FunctionTok{sort}\NormalTok{(}\AttributeTok{decreasing =} \ConstantTok{TRUE}\NormalTok{)}
\end{Highlighting}
\end{Shaded}

\captionof{chunk}{Using the pipe and \texttt{tapply()} to count applicants and compute the average loan amount from the loans data, sorted by average loan amount.}

\label{tapply-tidy-pipe} \linespread{2}
\vspace{3mm}\setlength{\parindent}{15pt}

\hypertarget{data-visualization}{%
\subsection{Data visualization}\label{data-visualization}}

The process of teaching and learning data visualization is challenging
\citep{nolaperr_2016}. Students are asked to engage with multiple
interrelated steps: (1) the conceptualization and design of the
visualization, (2) the translation of the design into the specific
syntax of the plotting tool or library, and (3) the cleaning,
conversion, and transformation of the data that will be represented in
the visualization. It is for the latter two tasks we believe the
tidyverse substantially improves on base R's functionality, particularly
when it comes to new learners.

Base R plotting is very powerful and flexible, but that flexibility
leads to idiosyncratic behavior and confusion. Take, for example, the
creation of a basic scatterplot of the variables \texttt{x} and
\texttt{y} stored in a data frame \texttt{d}. All of the following would
produce the same plot: \texttt{plot(d)},
\texttt{plot(y\textasciitilde{}x,\ d)}, \texttt{plot(d\$x,\ d\$y)},
\texttt{with(d,\ plot(x,y))}, and so on. This flexibility, made possible
by the S3 object system, is useful but can be overwhelming for new
learners. The difficulty is particularly apparent when students get
stuck and search for terms like ``scatterplot in R'' and come across
solutions which use an approach that differs substantially from their
instructor's preferred method (e.g., using the formula method when they
have only been shown \texttt{\$} for column access).

The issue of student confusion caused by idiosyncratic behavior gets
worse as needs expand beyond the most basic plotting primitives.
Students can easily \emph{use} custom plotting methods (e.g.,
\texttt{density(x)}), functions for more specific plotting primitives
(e.g., \texttt{boxplot(y\textasciitilde{}x)}), and methods combining
both (e.g., \texttt{hist(x)}) without building a higher-level
understanding of how to create a complex plot. That is, while all the
plotting tools are usable, it is hard for students to develop a mental
model of their commonalities without first having a deeper understanding
of R as a programming language, specifically around generic functions,
basic data structures, and classes.

As mentioned previously, most of base R's plotting functionality is
built around S3 specializations of \texttt{plot()} and similar
high-level plotting functions. In some cases, the plot types are
relatively easy to identify by name (e.g., \texttt{hist()},
\texttt{barplot()}, \texttt{boxplot()}) while others are less obvious
(e.g., \texttt{abline()}, \texttt{image()}, or \texttt{par()}). The base
R plotting functions are part of the \textbf{graphics} package, which is
loaded automatically by all R sessions.

To visualize using tidyverse principles, we use one of the core
tidyverse packages: ggplot2. In ggplot2, different plot types are
implemented using geometry functions (prefixed with \texttt{geom\_})
which map variables in the data to various aesthetic properties (e.g.,
horizontal and vertical position, size, color, etc.) of the plots. One
of the immediate advantages of using \texttt{geom\_*()} functions for
plotting is the reduction in the search space of possible plotting
functions. The prefix works effectively to organize the search space to
the possible functions and their associated documentation. The narrowing
is true whether within an IDE (using tab-completion), in the
documentation (looking just in the \texttt{g} section), or searching
online. In contrast, standard base graphics begin with a multitude of
different characters, so there is no obvious way to simply see all the
possibilities.

Additionally, the geometries implemented in ggplot2 use the same core
function arguments and share common aesthetics, which makes it easier to
pick up and explore new geometries, as well as quickly swap between
related visualization methods. For example, changing between boxplots
and violin plots (an augmented form of boxplots) only requires changing
\texttt{geom\_boxplot()} to \texttt{geom\_violin()}---the arguments to
the two functions remain the same.

As ggplot2 is built around the principles of the grammar of graphics
\citep{wilkinson2012grammar}, its syntax is designed to reflect the
process of building a visualization through the composition of layers
using the \texttt{+} operator. In contrast, base R graphics are
generally built up using multiple function calls (e.g.,
\texttt{plot(...,\ add=TRUE)} and \texttt{abline()}).

\ref{ggplot-exref} presents an example of side-by-side boxplots of
\texttt{loan\_amount} as a function of \texttt{application\_type} and
\texttt{homeownership}.

\linespread{1}

\begin{Shaded}
\begin{Highlighting}[]
\NormalTok{loans }\SpecialCharTok{|\textgreater{}}
  \FunctionTok{ggplot}\NormalTok{(}\FunctionTok{aes}\NormalTok{(}\AttributeTok{y =}\NormalTok{ loan\_amount, }\AttributeTok{x =}\NormalTok{ application\_type)) }\SpecialCharTok{+}
  \FunctionTok{geom\_boxplot}\NormalTok{() }\SpecialCharTok{+}
  \FunctionTok{facet\_wrap}\NormalTok{(}\SpecialCharTok{\textasciitilde{}}\NormalTok{ homeownership)}
\end{Highlighting}
\end{Shaded}

\begin{center}\includegraphics[width=1\linewidth]{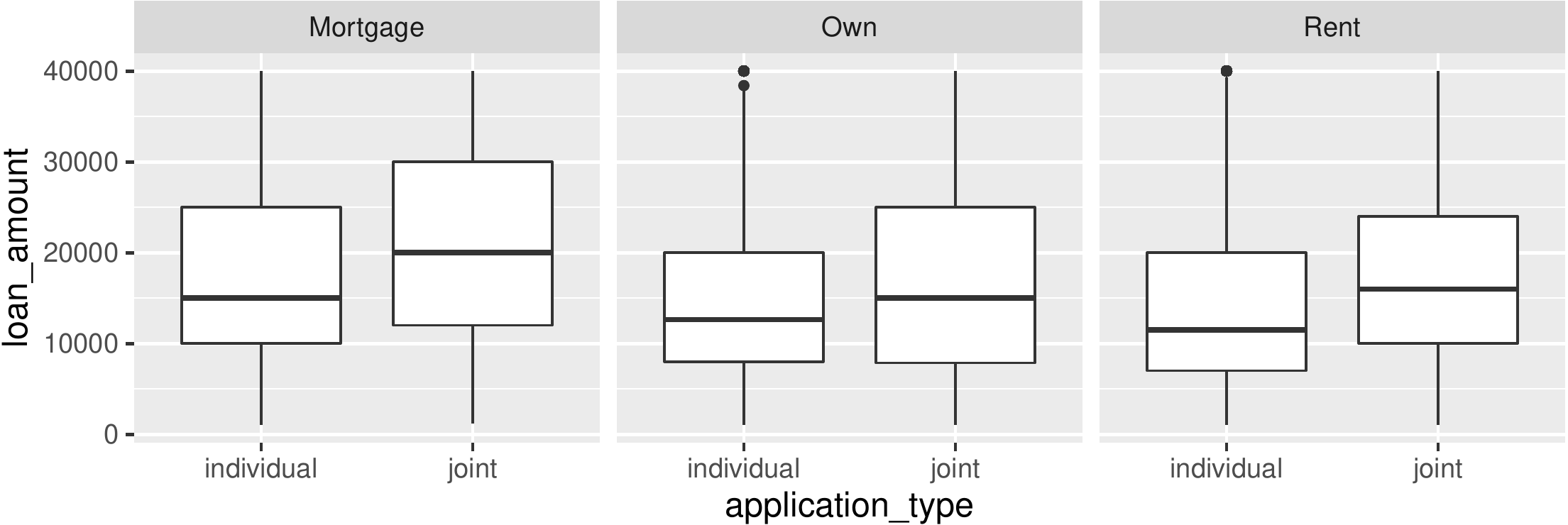} \end{center}
\captionof{chunk}{Using ggplot2 to create boxplots of \texttt{loan\_amount} broken down by both \texttt{application\_type} and \texttt{homeownership}.}

\label{ggplot-exref} \linespread{2}
\vspace{3mm}\setlength{\parindent}{15pt}

The figure is constructed in three calls: one to bind the data and map
the variables to aesthetics, another to draw the boxplot, and the last
to facet by levels of \texttt{homeownership}. A similar plot can be
constructed using base R graphics (see \ref{base-plot-ex}), however the
process of creating facets is more burdensome as it requires using
creation of separate plots with a \texttt{for()} loop and subsetting to
iterate over the unique levels of \texttt{homeownership}.

\linespread{1}

\begin{Shaded}
\begin{Highlighting}[]
\NormalTok{levels }\OtherTok{\textless{}{-}} \FunctionTok{sort}\NormalTok{(}\FunctionTok{unique}\NormalTok{(loans}\SpecialCharTok{$}\NormalTok{homeownership))}
\NormalTok{n }\OtherTok{\textless{}{-}} \FunctionTok{length}\NormalTok{(levels)}

\FunctionTok{par}\NormalTok{(}\AttributeTok{mfrow =} \FunctionTok{c}\NormalTok{(}\DecValTok{1}\NormalTok{,n))}
\ControlFlowTok{for}\NormalTok{(i }\ControlFlowTok{in} \FunctionTok{seq\_len}\NormalTok{(n)) \{}
  \FunctionTok{boxplot}\NormalTok{(}
\NormalTok{    loan\_amount }\SpecialCharTok{\textasciitilde{}}\NormalTok{ application\_type, }
    \AttributeTok{data =}\NormalTok{ loans[loans}\SpecialCharTok{$}\NormalTok{homeownership }\SpecialCharTok{==}\NormalTok{ levels[i],],}
    \AttributeTok{main =}\NormalTok{ levels[i]}
\NormalTok{  )}
\NormalTok{\}}
\end{Highlighting}
\end{Shaded}

\begin{center}\includegraphics[width=1\linewidth]{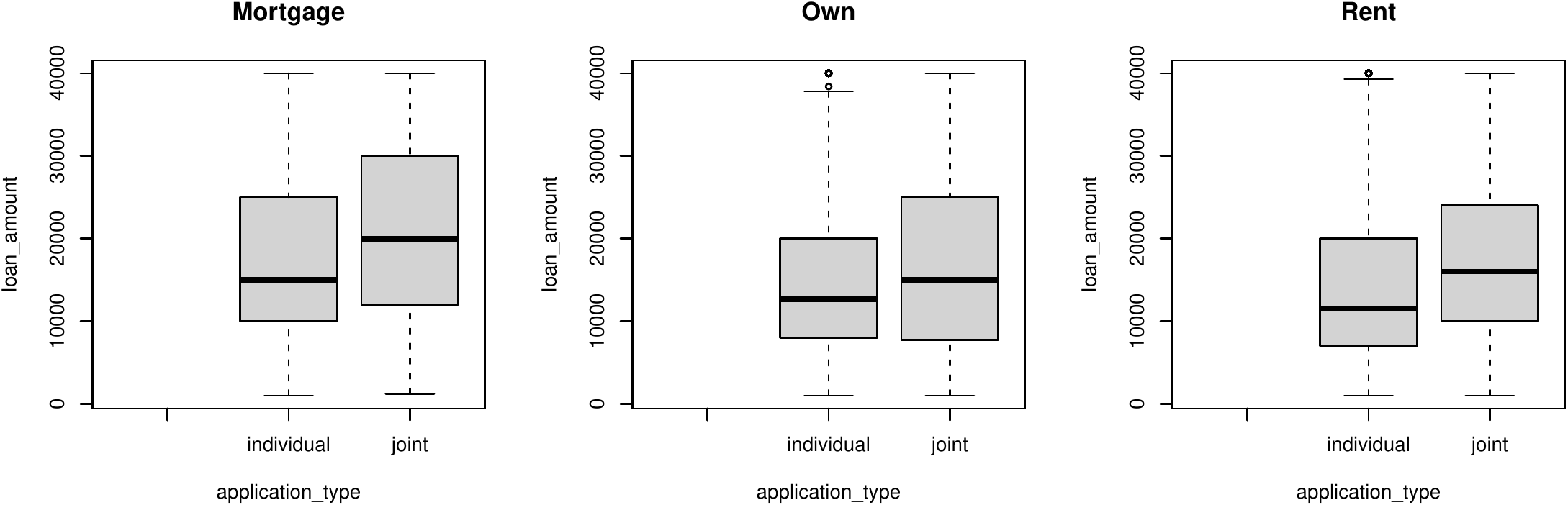} \end{center}
\captionof{chunk}{Using base R to create boxplots of \texttt{loan\_amount} broken down by both \texttt{application\_type} and \texttt{homeownership}.}

\label{base-plot-ex} \linespread{2}
\vspace{3mm}\setlength{\parindent}{15pt}

As seen in \ref{ggplot-ex2}, in ggplot2 it is straightforward to extend
the original boxplot to more data dimensions by mapping another variable
to an additional aesthetic. For example, adding
\texttt{fill\ =\ bankruptcy} to the \texttt{aes()} call will create a
plot that now displays four dimensions of the original data.

\linespread{1}

\begin{Shaded}
\begin{Highlighting}[]
\NormalTok{loans }\SpecialCharTok{|\textgreater{}}
  \FunctionTok{ggplot}\NormalTok{(}\FunctionTok{aes}\NormalTok{(}\AttributeTok{y =}\NormalTok{ loan\_amount, }\AttributeTok{x =}\NormalTok{ application\_type, }\AttributeTok{fill =}\NormalTok{ bankruptcy)) }\SpecialCharTok{+}
  \FunctionTok{geom\_boxplot}\NormalTok{() }\SpecialCharTok{+}
  \FunctionTok{facet\_wrap}\NormalTok{(}\SpecialCharTok{\textasciitilde{}}\NormalTok{ homeownership)}
\end{Highlighting}
\end{Shaded}

\begin{center}\includegraphics[width=1\linewidth]{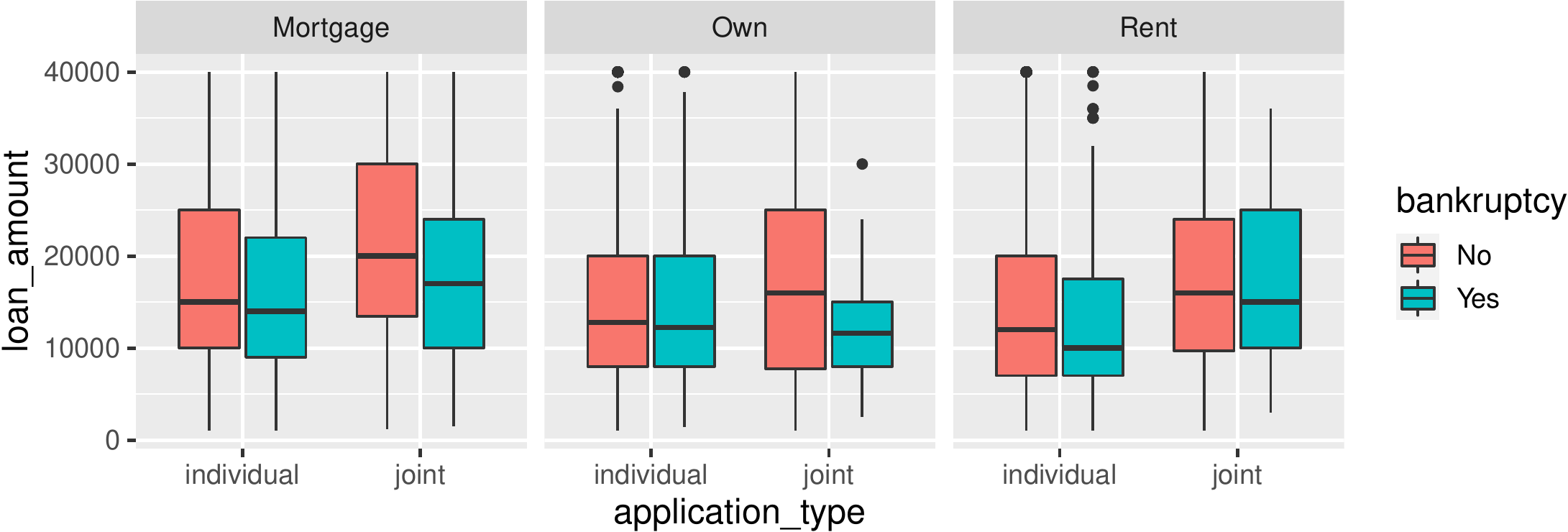} \end{center}
\captionof{chunk}{Using ggplot2 to create boxplots of \texttt{loan\_amount} broken down by both \texttt{application\_type} and \texttt{homeownership}, filled by \texttt{bankruptcy}.}

\label{ggplot-ex2} \linespread{2}
\vspace{3mm}\setlength{\parindent}{15pt}

Splitting the data to create multiple boxplots based on different
variables is possible with base R, however the implementation is more
complex and verbose, and it is left as an exercise to the reader.

Next we generate scatterplots of interest rate as a function of income,
where the points and linear fits are colored by the bankruptcy status of
the loan recipient.

\linespread{1}

\begin{Shaded}
\begin{Highlighting}[]
\NormalTok{loans }\SpecialCharTok{|\textgreater{}}
  \FunctionTok{ggplot}\NormalTok{(}\FunctionTok{aes}\NormalTok{(}\AttributeTok{y =}\NormalTok{ interest\_rate, }\AttributeTok{x =}\NormalTok{ annual\_income, }\AttributeTok{color =}\NormalTok{ bankruptcy)) }\SpecialCharTok{+}
  \FunctionTok{geom\_point}\NormalTok{(}\AttributeTok{alpha =} \FloatTok{0.1}\NormalTok{) }\SpecialCharTok{+} 
  \FunctionTok{geom\_smooth}\NormalTok{(}\AttributeTok{method =} \StringTok{"lm"}\NormalTok{, }\AttributeTok{size =} \DecValTok{2}\NormalTok{, }\AttributeTok{se =} \ConstantTok{FALSE}\NormalTok{) }\SpecialCharTok{+} 
  \FunctionTok{scale\_x\_log10}\NormalTok{(}\AttributeTok{labels =}\NormalTok{ scales}\SpecialCharTok{::}\FunctionTok{label\_dollar}\NormalTok{()) }\SpecialCharTok{+}
  \FunctionTok{scale\_y\_continuous}\NormalTok{(}\AttributeTok{labels =}\NormalTok{ scales}\SpecialCharTok{::}\FunctionTok{label\_percent}\NormalTok{(}\AttributeTok{scale =} \DecValTok{1}\NormalTok{)) }\SpecialCharTok{+}
  \FunctionTok{scale\_color\_manual}\NormalTok{(}\AttributeTok{values =} \FunctionTok{c}\NormalTok{(}\StringTok{"\#E69F00"}\NormalTok{, }\StringTok{"\#56B4E9"}\NormalTok{)) }\SpecialCharTok{+}
  \FunctionTok{labs}\NormalTok{(}
    \AttributeTok{x =} \StringTok{"Annual Income"}\NormalTok{, }
    \AttributeTok{y =} \StringTok{"Interest Rate"}\NormalTok{, }
    \AttributeTok{color =} \StringTok{"Previous}\SpecialCharTok{\textbackslash{}n}\StringTok{Bankruptcy"}
\NormalTok{    )}
\end{Highlighting}
\end{Shaded}

\begin{center}\includegraphics[width=0.8\linewidth]{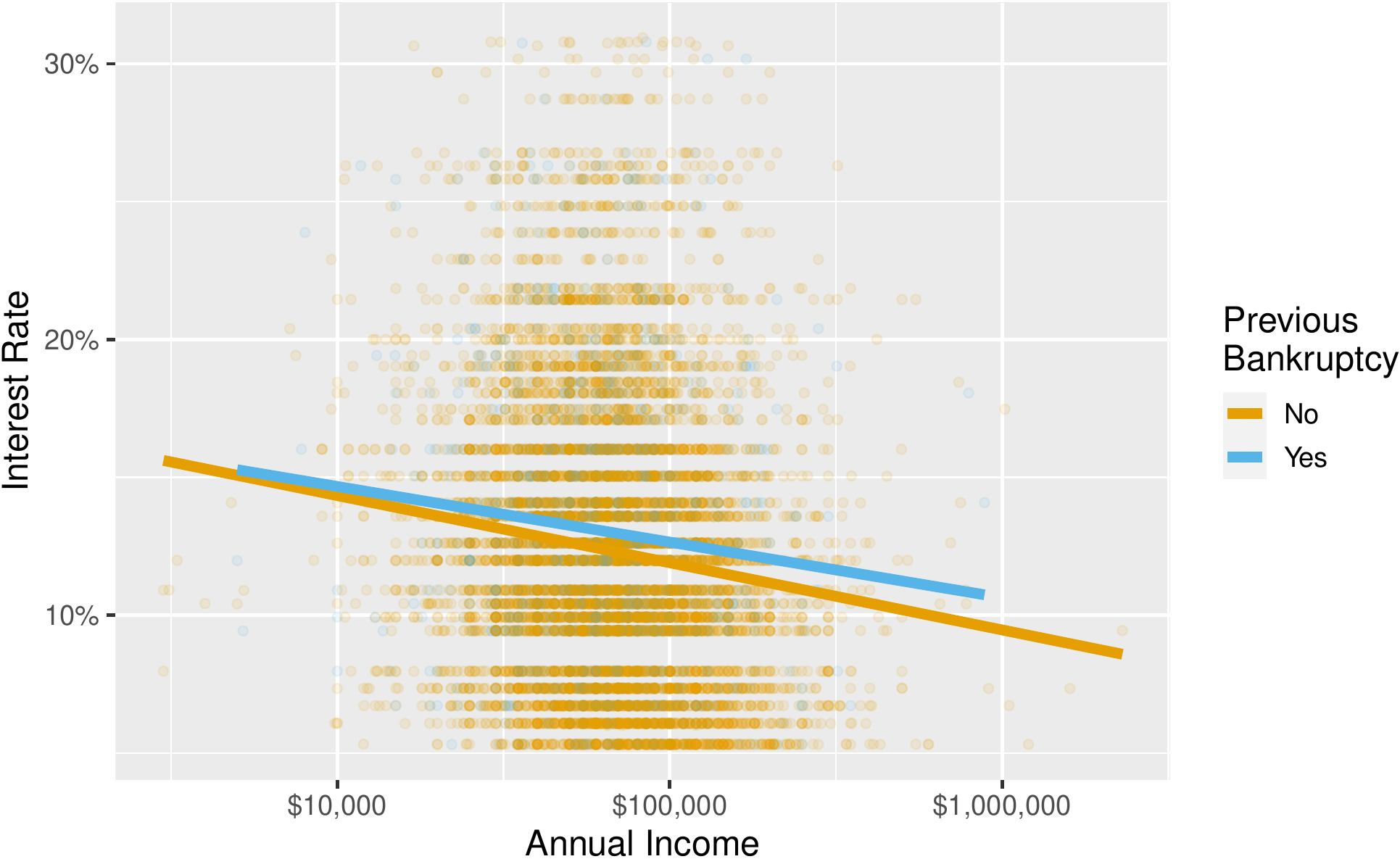} \end{center}
\captionof{chunk}{Using ggplot2 to create a scatterplot of \texttt{interest\_rate} versus annual income, colored by \texttt{bankruptcy}.}

\label{ggplot-scat} \linespread{2}
\vspace{3mm}\setlength{\parindent}{15pt}

We note that \ref{ggplot-scat} requires seven function calls, but each
call builds on a common framework and interface.

Base R code is used in \ref{base-scat} to create a similar scatterplot,
but the syntax requires careful study of the documentation for
\texttt{par()} and multiple separate calls of \texttt{lm()}.

\linespread{1}

\begin{Shaded}
\begin{Highlighting}[]
\NormalTok{cols }\OtherTok{=} \FunctionTok{c}\NormalTok{(}\AttributeTok{No =} \StringTok{"\#E69F00"}\NormalTok{, }\AttributeTok{Yes =} \StringTok{"\#56B4E9"}\NormalTok{)}

\FunctionTok{plot}\NormalTok{(}
\NormalTok{  loans}\SpecialCharTok{$}\NormalTok{annual\_income,}
\NormalTok{  loans}\SpecialCharTok{$}\NormalTok{interest\_rate,}
  \AttributeTok{pch =} \DecValTok{16}\NormalTok{,}
  \AttributeTok{col =} \FunctionTok{adjustcolor}\NormalTok{(cols[loans}\SpecialCharTok{$}\NormalTok{bankruptcy], }\AttributeTok{alpha.f =} \FloatTok{0.1}\NormalTok{),}
  \AttributeTok{log =} \StringTok{"x"}\NormalTok{,}
  \AttributeTok{xlab =} \StringTok{"Annual Income ($)"}\NormalTok{,}
  \AttributeTok{ylab =} \StringTok{"Interest Rate (\%)"}\NormalTok{,}
  \AttributeTok{xaxp =} \FunctionTok{c}\NormalTok{(}\DecValTok{1000}\NormalTok{, }\DecValTok{10000000}\NormalTok{, }\DecValTok{1}\NormalTok{)}
\NormalTok{)}

\NormalTok{lm\_b\_no }\OtherTok{=} \FunctionTok{lm}\NormalTok{(}
\NormalTok{  interest\_rate }\SpecialCharTok{\textasciitilde{}} \FunctionTok{log10}\NormalTok{(annual\_income), }
  \AttributeTok{data =}\NormalTok{ loans[loans}\SpecialCharTok{$}\NormalTok{bankruptcy }\SpecialCharTok{==} \StringTok{"No"}\NormalTok{,]}
\NormalTok{)}
\NormalTok{lm\_b\_yes }\OtherTok{=} \FunctionTok{lm}\NormalTok{(}
\NormalTok{  interest\_rate }\SpecialCharTok{\textasciitilde{}} \FunctionTok{log10}\NormalTok{(annual\_income), }
  \AttributeTok{data =}\NormalTok{ loans[loans}\SpecialCharTok{$}\NormalTok{bankruptcy }\SpecialCharTok{==} \StringTok{"Yes"}\NormalTok{,]}
\NormalTok{)}

\FunctionTok{abline}\NormalTok{(lm\_b\_no, }\AttributeTok{col =}\NormalTok{ cols[}\StringTok{"No"}\NormalTok{], }\AttributeTok{lwd =} \DecValTok{3}\NormalTok{)}
\FunctionTok{abline}\NormalTok{(lm\_b\_yes, }\AttributeTok{col =}\NormalTok{ cols[}\StringTok{"Yes"}\NormalTok{], }\AttributeTok{lwd =} \DecValTok{3}\NormalTok{)}

\FunctionTok{legend}\NormalTok{(}
  \StringTok{"topright"}\NormalTok{, }
  \AttributeTok{legend =} \FunctionTok{c}\NormalTok{(}\StringTok{"Yes"}\NormalTok{, }\StringTok{"No"}\NormalTok{), }
  \AttributeTok{title =} \StringTok{"Previous}\SpecialCharTok{\textbackslash{}n}\StringTok{Bankruptcy"}\NormalTok{, }
  \AttributeTok{col =}\NormalTok{ cols[}\FunctionTok{c}\NormalTok{(}\StringTok{"Yes"}\NormalTok{, }\StringTok{"No"}\NormalTok{)], }
  \AttributeTok{pch =} \DecValTok{16}\NormalTok{, }\AttributeTok{lwd =} \DecValTok{1}
\NormalTok{)}
\end{Highlighting}
\end{Shaded}

\begin{center}\includegraphics[width=0.8\linewidth]{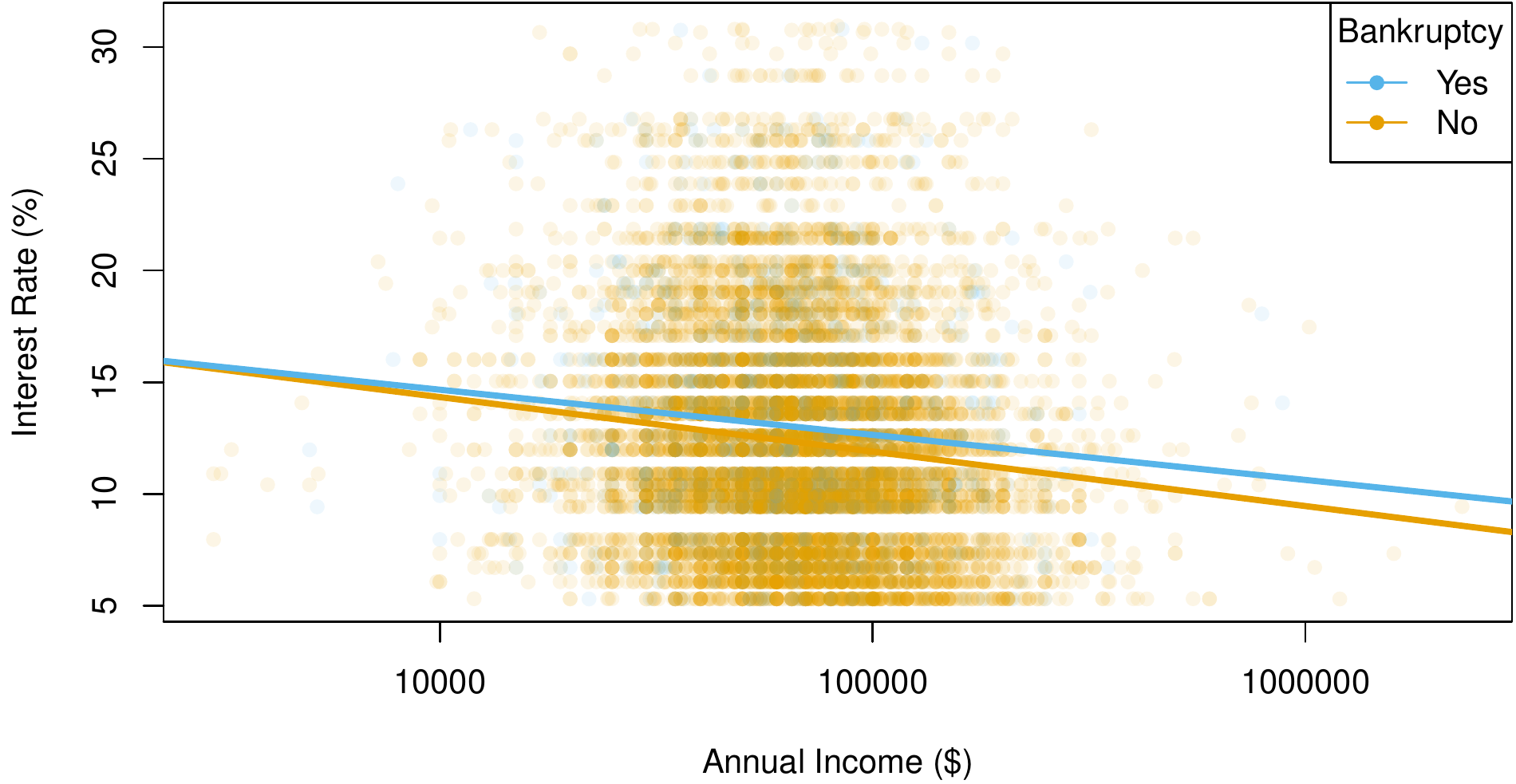} \end{center}
\captionof{chunk}{Using base R to create a scatterplot of \texttt{interest\_rate} versus annual income, colored by \texttt{bankruptcy}.}

\label{base-scat} \linespread{2}
\vspace{3mm}\setlength{\parindent}{15pt}

We note several differences between the two scatterplot implementations.
The ggplot2 approach uses function arguments vs.~base graphics
parameters. Some of the parameters (e.g., \texttt{xaxp}, \texttt{log})
are idiosyncratic and difficult to find in the documentation. Specifying
colors and alpha transparency levels are quite different in the two
implementations. In base graphics, legends require a separate function
with additional manual bookkeeping, which increases the potential for
human error. Explicit iteration is required to create multivariate plots
and the user is required to specify the display structure. Care is
needed in constructing axis values.

We note that the \texttt{+} operator in the ggplot2 syntax differs from
the pipe (\texttt{\textbar{}\textgreater{}}) operator for historical
reasons \citep{wickham2015pipe}. We have found, however, that the
\texttt{+} operator is straightforward for students to learn. The
\texttt{+} operator is consistent with the layering aspect of creating a
plot. Additionally, the error message when
\texttt{\textbar{}\textgreater{}} is used in place of \texttt{+}
explicitly asks:
\texttt{Did\ you\ use\ \%\textgreater{}\%\ instead\ of\ +?}.\footnote{As
  of this writing, the ggplot2 error message still references the
  magrittr pipe (\texttt{\%\textgreater{}\%}), even when you use the
  native pipe (\texttt{\textbar{}\textgreater{}}).}

\hypertarget{sec:extra}{%
\subsection{Beyond wrangling, statistical summaries, and
visualizations}\label{sec:extra}}

While data wrangling and data visualization are high on the list of
important tasks for working with data, there are many additional tasks
and tools needed for a complete data analysis. We argue that setting
your students up to understand the tidyverse approach will not only help
them work fluently with data at the exploratory data analysis stage, but
will also provide a solid grounding for inference, modeling, working
with databases (e.g., SQL), and other data-focused operations.

\textbf{Statistical inference} can be taught using the \textbf{infer}
package \citep{R-infer}, which uses a consistent syntax for one- and
two-sample inferential techniques for tidy data. The package implements
both computational methods like randomization tests and bootstrapping as
well as mathematical models like t- and z-tests. With the infer syntax,
students learn a single set of functions that walk through the
inferential process and focus attention on each step of the process.
Indeed, the functions themselves are named to reinforce the conceptual
understanding of the process. For example, a student will
\texttt{specify()} the variables \emph{and then} \texttt{hypothesize()}
about the conditions \emph{and then} \texttt{generate()} a sampling
distribution \emph{and then} \texttt{calculate()} the statistic of
interest. The output of these functions can be wrangled (e.g., to
determine a p-value) or visualized (e.g., to inspect the null sampling
distribution).

When doing \textbf{modeling and machine learning}, many data analysts
rely on functions like \texttt{lm()} and \texttt{glm()} as important
tools in the analysis process. To avoid reinventing the wheel, packages
like \textbf{broom} \citep{R-broom} allow \textbf{tidy modeling with
base R functions} to extend the convention of data frame as input and
data frame as output for modeling tasks. Using \texttt{lm()} as an
example, there are model output pieces that: 1) describe each of the
variables (e.g., the coefficients); 2) describe each of the
observational units (e.g., the residuals); and 3) describe the entire
model (e.g., \(R^2\)). Since the \texttt{lm()} output is a list with a
specialized print method, it is not immediately clear how to access
certain components of it (e.g., the value of the slope or the intercept)
programatically. The broom paradigm---which can be applied to
\texttt{lm()} and \texttt{glm()} as well as many other model
objects---uses \texttt{tidy()} for output which describes variables,
\texttt{augment()} to describe observational units, and
\texttt{glance()} to describe the entire model. The outputs of each of
these functions is a tibble, which makes it easy to extract values like
coefficient estimates or p-values using data wrangling functions offered
by dplyr. This, in turn, makes it straightforward to include these
extracted values within text in computational documents (e.g., R
Markdown documents) using inline code chunks, hence easing reproducible
communication of results.

In addition to modeling done using base R functions, a educator or
student interested in machine learning methods or approaches (e.g.,
cross validation) may benefit from \textbf{tidy modeling functionality}
available in the \textbf{tidymodels} package \citep{R-tidymodels}. The
core idea of tidymodels is to simplify the practice of modeling by
pre-processing \emph{and then} training \emph{and then} validating
models. The pre-processing step can scale variables or filter for highly
correlated covariates. The training step can use anything from a linear
model to a support vector machine to a neural network. The validating
step incorporates cross validation or prediction on a test dataset to
evaluate the user's metric of choice.

\textbf{Database technologies} are supported using the dplyr
\citep{R-dplyr} and \textbf{dbplyr} \citep{R-dbplyr} packages, which
facilitate access to SQL databases using the same general syntax and
idioms learned with the tidyverse. As an example, we can access and
summarize data from a publicly accessible repository of audiological
measurements \citep{voss_2019} using the same form as our earlier
wrangling (see Section \ref{sec:wrangle}).

\linespread{1}

\begin{Shaded}
\begin{Highlighting}[]
\NormalTok{db }\OtherTok{\textless{}{-}}\NormalTok{ DBI}\SpecialCharTok{::}\FunctionTok{dbConnect}\NormalTok{(}
\NormalTok{  RMySQL}\SpecialCharTok{::}\FunctionTok{MySQL}\NormalTok{(),}
  \AttributeTok{dbname =} \StringTok{"wai"}\NormalTok{, }
  \AttributeTok{host =} \StringTok{"scidb.smith.edu"}\NormalTok{, }
  \AttributeTok{username =} \StringTok{"waiuser"}\NormalTok{, }
  \AttributeTok{password =} \StringTok{"smith\_waiDB"}  \CommentTok{\# publicly accessible database}
\NormalTok{)}

\NormalTok{db }\SpecialCharTok{|\textgreater{}}
  \FunctionTok{tbl}\NormalTok{(}\StringTok{"Subjects"}\NormalTok{) }\SpecialCharTok{|\textgreater{}} 
  \FunctionTok{group\_by}\NormalTok{(AgeCategoryFirstMeasurement) }\SpecialCharTok{|\textgreater{}}
  \FunctionTok{summarize}\NormalTok{(}\AttributeTok{num\_people =} \FunctionTok{n}\NormalTok{())}
\end{Highlighting}
\end{Shaded}

\begin{verbatim}
## # Source:   lazy query [?? x 2]
## # Database: mysql 8.0.27-0ubuntu0.20.04.1 [@scidb.smith.edu:/wai]
##   AgeCategoryFirstMeasurement num_people
##   <chr>                            <dbl>
## 1 Adult                              775
## 2 Infant                            2215
## 3 Child                              116
## 4 NICU                                41
\end{verbatim}

\captionof{chunk}{Applying tidyverse wrangling to data which has been queried from a SQL database.}

\label{SQL-ex} \linespread{2} \vspace{3mm}\setlength{\parindent}{15pt}

Note that the output of the wrangled SQL data is a \texttt{tibble}
similar to that of Section \ref{sec:wrangle}, but with additional
information about the MySQL database server.

\hypertarget{sec:teach}{%
\section{Pedagogical strengths of the tidyverse}\label{sec:teach}}

In this section, we expand on the core benefit of the tidyverse outlined
in Section \ref{sec:tidyverse}, consistency, by highlighting pedagogical
strengths with respect to mixability, scalability, user-centered design,
readability, community, and shared syntax.

\hypertarget{sec:mixability}{%
\subsection{Mixability}\label{sec:mixability}}

Consistent syntax and interface are hallmarks of the tidyverse's design
principles (as outlined in Section \ref{sec:principles}). As
instructors, we strive for consistency in how we use the tidyverse in
our teaching. To achieve this, we avoid mixing-and-matching tidyverse
patterns with base R patterns. An example of inconsistent behavior to
avoid is using base graphics for boxplots and ggplot2 for scatterplots
or using \texttt{dplyr::count()} for creating a frequency table for two
categorical variables but then using a function from the
\texttt{apply()} family for creating summaries for a numerical variable
across levels of a categorical variable. However, it's not possible to
write ``tidyverse code'' without using base R functions and this
mixability is a strength of the tidyverse, allowing students to learn
many base R functions (e.g., \texttt{mean()}, \texttt{sd()},
\texttt{quantile()}, \texttt{dnorm()}) while learning the tidyverse
framework. Additionally, because the input and output of tidyverse
functions are ``normal'' R objects (typically, a data frame), an
instructor can be consistent without coding exclusively in the
tidyverse.

In Section \ref{sec:extra} we illustrate how the infer package can be
used in an introductory statistics course to extend the tidyverse
framework to include statistical inference. However, one can certainly
teach introductory statistics by combining functionality from the most
popular tidyverse functionality (i.e., dplyr and ggplot2) for wrangling
with base R implementations of inferential functions (e.g.,
\texttt{t.test()}, \texttt{chisq.test()}, etc.). The combination of
functions will necessitate some inconsistency, since some base R
inferential functions accept vectors as inputs rather than data frames.
However, as with explaining the difference between the \texttt{+} and
\texttt{\textbar{}\textgreater{}}, instructors can explain that data
wrangling and visualization always use data frames, but that inferential
functions sometimes use vectors. We believe that inference is smoother
with the infer package, but a decision not to adopt infer should not
preclude an instructor from adopting the tidyverse for other tasks.

A similar argument holds for statistical modeling, where \texttt{lm()}
can be used alongside tidyverse code. We contend that broom and/or
tidymodels reduce friction when analyzing data that can arise due to
inconsistent input and output types of base R's modeling functions, but
several of the authors still teach \texttt{lm()} in their own courses.

\hypertarget{sec:scalability}{%
\subsection{Scalability}\label{sec:scalability}}

In Section \ref{sec:wrangle}, we illustrated how the
\texttt{group\_by()} and \texttt{summarize()} verbs make it easy to
``roll up'' a data frame by groups. Next, we consider extending that
analysis in two different ways: by adding one or more summary
computations; and by adding one or more additional grouping variables.
In both cases, we argue that tidyverse code adapts more easily than base
R by not requiring a student to learn any additional functions.

First, in Example \ref{tidy-summary-again}, we note that adding an
aggregate computation involves only a comma and the expression involving
the relevant summary function in the call to \texttt{summarize()}. Here,
we compute two quantities: a count (\texttt{num\_applicants}) and a mean
(\texttt{avg\_loan\_amount}).

\linespread{1}

\begin{Shaded}
\begin{Highlighting}[]
\NormalTok{loans }\SpecialCharTok{|\textgreater{}}
  \FunctionTok{group\_by}\NormalTok{(homeownership) }\SpecialCharTok{|\textgreater{}}
  \FunctionTok{summarize}\NormalTok{(}
    \AttributeTok{num\_applicants =} \FunctionTok{n}\NormalTok{(),}
    \AttributeTok{avg\_loan\_amount =} \FunctionTok{mean}\NormalTok{(loan\_amount)}
\NormalTok{  ) }\SpecialCharTok{|\textgreater{}}
  \FunctionTok{arrange}\NormalTok{(}\FunctionTok{desc}\NormalTok{(avg\_loan\_amount))}
\end{Highlighting}
\end{Shaded}

\begin{verbatim}
## # A tibble: 3 x 3
##   homeownership num_applicants avg_loan_amount
##   <chr>                  <int>           <dbl>
## 1 Mortgage                4778          18132.
## 2 Own                     1350          15665.
## 3 Rent                    3848          14396.
\end{verbatim}

\captionof{chunk}{Using the tidyverse to count applicants and compute the average loan amount from the loans data, sorted by average loan amount.}

\label{tidy-summary-again} \linespread{2}
\vspace{3mm}\setlength{\parindent}{15pt}

If we wanted to compute \(p\) quantities, for any integer \(p > 1\), it
is straightforward to add \(p\) arguments to \texttt{summarize()}. Thus,
this type of operation is \emph{scalable}, because the number of lines
of code is proportional to the number of quantities computed.

Additionally, if the variables to be summarized can be selected based on
their types or names, we can use \texttt{dplyr::across()} to summarize
many variables with one line of code. In example \ref{across}, we show
how to calculate the mean for across variables in the \texttt{loans}
dataset that contain the character string \texttt{"paid"} for each
\texttt{homeownership} group.

\linespread{1}

\begin{Shaded}
\begin{Highlighting}[]
\NormalTok{loans }\SpecialCharTok{|\textgreater{}} 
  \FunctionTok{group\_by}\NormalTok{(homeownership) }\SpecialCharTok{|\textgreater{}} 
  \FunctionTok{summarise}\NormalTok{(}\FunctionTok{across}\NormalTok{(}\FunctionTok{contains}\NormalTok{(}\StringTok{"paid"}\NormalTok{), mean))}
\end{Highlighting}
\end{Shaded}

\begin{verbatim}
## # A tibble: 3 x 5
##   homeownership paid_total paid_principal paid_interest paid_late_fees
##   <chr>              <dbl>          <dbl>         <dbl>          <dbl>
## 1 Mortgage           2734.          2089.          646.         0.0810
## 2 Own                2512.          1950.          563.         0.111 
## 3 Rent               2191.          1635.          555.         0.171
\end{verbatim}

\captionof{chunk}{Using the tidyverse to calculate mean amounts across variables that contain the character string 'paid' for each homeownership group.}

\label{across} \linespread{2} \vspace{3mm}\setlength{\parindent}{15pt}

Achieving scalability in base R is possible, but the approaches involve
additional programming concepts. A conceptually straightforward base R
approach to scaling up is to call \texttt{tapply()} \(p\) times and
combine the resulting vectors using \texttt{cbind()} (see
\ref{base-scaling-1}). While the approach also involves \(O(p)\) lines
of code, you would need to type the names of the two variables
(\texttt{loan\_amount} and \texttt{homeownership}) \(p\) times, instead
of once.

\linespread{1}

\begin{Shaded}
\begin{Highlighting}[]
\FunctionTok{with}\NormalTok{(}
\NormalTok{  loans, }
  \FunctionTok{cbind}\NormalTok{(}
    \AttributeTok{num\_applicants =} \FunctionTok{tapply}\NormalTok{(loan\_amount, homeownership, length),}
    \AttributeTok{avg\_loan\_amount =} \FunctionTok{tapply}\NormalTok{(loan\_amount, homeownership, mean)}
\NormalTok{  )}
\NormalTok{)}
\end{Highlighting}
\end{Shaded}

\begin{verbatim}
##          num_applicants avg_loan_amount
## Mortgage           4778        18132.45
## Own                1350        15665.44
## Rent               3848        14396.44
\end{verbatim}

\captionof{chunk}{Using \texttt{tapply()} in base R to calculate multiple summary statistics.}

\label{base-scaling-1} \linespread{2}
\vspace{3mm}\setlength{\parindent}{15pt}

A base R construction that is more scalable involves writing a custom
summary function and iteratively combining the results using
\texttt{do.call()} and \texttt{rbind()} (see \ref{base-scaling-2}). The
approach requires two programming concepts (writing a user-defined
function and iterating a function) that are likely beyond the scope of
introductory statistics or data science.

\linespread{1}

\begin{Shaded}
\begin{Highlighting}[]
\NormalTok{my\_summary }\OtherTok{\textless{}{-}} \ControlFlowTok{function}\NormalTok{(x) \{ }
  \FunctionTok{data.frame}\NormalTok{(}
    \AttributeTok{num\_applicants =} \FunctionTok{length}\NormalTok{(x),}
    \AttributeTok{avg\_loan\_amount =} \FunctionTok{mean}\NormalTok{(x)}
\NormalTok{  )}
\NormalTok{\}}
\FunctionTok{do.call}\NormalTok{(rbind, }\FunctionTok{with}\NormalTok{(loans, }\FunctionTok{tapply}\NormalTok{(loan\_amount, homeownership, my\_summary)))}
\end{Highlighting}
\end{Shaded}

\begin{verbatim}
##          num_applicants avg_loan_amount
## Mortgage           4778        18132.45
## Own                1350        15665.44
## Rent               3848        14396.44
\end{verbatim}

\captionof{chunk}{Using a custom function in base R to calculate multiple summary statistics.}

\label{base-scaling-2} \linespread{2}
\vspace{3mm}\setlength{\parindent}{15pt}

Second, we note that adding a second grouping variable in the tidyverse
involves only adding another argument to \texttt{group\_by()}. Moreover,
adding \(k\) grouping variables involves adding \(k\) items to
\texttt{group\_by()} (see \ref{tidy-scaling-2}). No additional
programming knowledge is necessary.

\linespread{1}

\begin{Shaded}
\begin{Highlighting}[]
\NormalTok{loans }\SpecialCharTok{|\textgreater{}}
  \FunctionTok{group\_by}\NormalTok{(homeownership, verified\_income) }\SpecialCharTok{|\textgreater{}}
  \FunctionTok{summarize}\NormalTok{(}
    \AttributeTok{num\_applicants =} \FunctionTok{n}\NormalTok{(),}
    \AttributeTok{avg\_loan\_amount =} \FunctionTok{mean}\NormalTok{(loan\_amount)}
\NormalTok{  )}
\end{Highlighting}
\end{Shaded}

\begin{verbatim}
## # A tibble: 9 x 4
## # Groups:   homeownership [3]
##   homeownership verified_income num_applicants avg_loan_amount
##   <chr>         <fct>                    <int>           <dbl>
## 1 Mortgage      Not Verified              1580          14739.
## 2 Mortgage      Source Verified           1963          19085.
## 3 Mortgage      Verified                  1235          20960.
## 4 Own           Not Verified               495          13161.
## 5 Own           Source Verified            547          16537.
## 6 Own           Verified                   308          18142.
## 7 Rent          Not Verified              1498          12287.
## 8 Rent          Source Verified           1605          14964.
## 9 Rent          Verified                   745          17413.
\end{verbatim}

\captionof{chunk}{Using the tidyverse to add a layer of grouping before calculating summary statistics.}

\label{tidy-scaling-2} \linespread{2}
\vspace{3mm}\setlength{\parindent}{15pt}

In base R, additional grouping variables can be added by wrapping the
set of grouping variables names in \texttt{list()}. Here again, while
the approach is scalable, it introduces a data structure (\texttt{list})
that is not typically necessary for introductory classes (see
\ref{base-scaling-3}). The output also omits the factor levels that
correspond to each row, which introduces potential confusion.

\linespread{1}

\begin{Shaded}
\begin{Highlighting}[]
\FunctionTok{do.call}\NormalTok{(}
\NormalTok{  rbind, }
  \FunctionTok{with}\NormalTok{(}
\NormalTok{    loans, }
    \FunctionTok{tapply}\NormalTok{(loan\_amount, }\FunctionTok{list}\NormalTok{(homeownership, verified\_income), my\_summary)}
\NormalTok{  )}
\NormalTok{)}
\end{Highlighting}
\end{Shaded}

\begin{verbatim}
##   num_applicants avg_loan_amount
## 1           1580        14739.15
## 2            495        13161.21
## 3           1498        12287.43
## 4           1963        19084.74
## 5            547        16536.97
## 6           1605        14964.45
## 7           1235        20960.04
## 8            308        18142.29
## 9            745        17413.39
\end{verbatim}

\captionof{chunk}{Using base R to add a layer of grouping before calculating summary statistics.}

\label{base-scaling-3} \linespread{2}
\vspace{3mm}\setlength{\parindent}{15pt}

Thus, in both cases, while scalable programming in base R is possible,
it brings with it additional extraneous programming concepts that may
distract---rather than support---students from learning statistics and
data science. Some of the confusion is because base R functions operate
on vectors, matrices, and data frames idiosyncratically. In contrast,
because tidyverse functions focus on data frames, users of the tidyverse
may be able to fall into the ``pit of success,'' \citep{wickham2016user}
wherein scaling one's analysis is natural and minimizes extraneous
cognition or bookkeeping.

\hypertarget{sec:user-centered}{%
\subsection{User-centered design}\label{sec:user-centered}}

The tidyverse has been developed with a user-centered design process
\citep{kling1977organizational, norman1986}, that can also be considered
learner-centered \citep{solowayetal1994}. While many R packages are
designed once and then updated incrementally with bug fixes, a number of
the packages within the tidyverse have undergone large scale API changes
to improve usability.

\begin{figure}[H]

{\centering \includegraphics[width=0.95\linewidth]{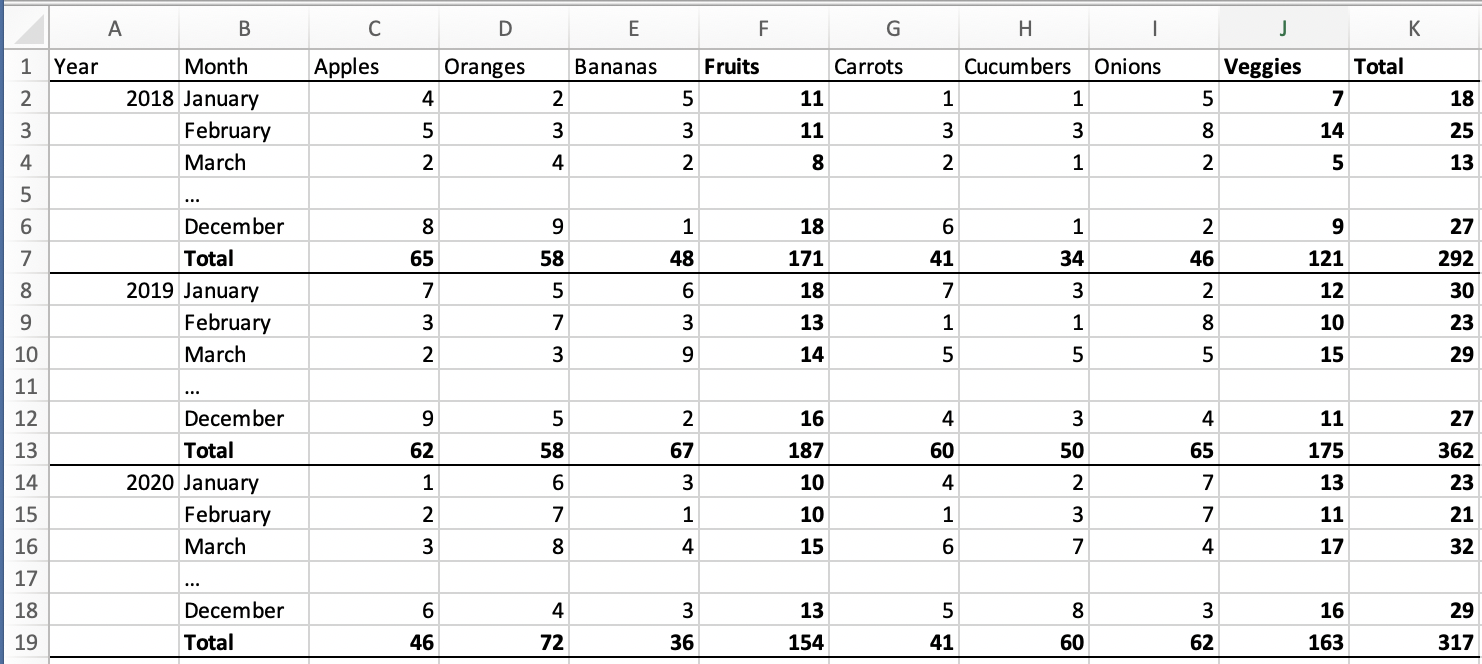} 

}

\caption{Non-tidy data which can be wrangled using \texttt{pivot\_longer()}.}\label{fig:fruitveggie}
\end{figure}

A prime example of the user-centered approach to development in the
tidyverse is the evolution of the functions for reshaping data.
Reshaping data is a key data wrangling skill, as data does not always
come in a format conducive to analysis. For example, consider data
containing counts of fruits and vegetables sold at a produce stand where
rows are years and months (and totals!) and columns are fruits and
vegetables (and totals!), as in Figure \ref{fig:fruitveggie}. Note that
in the sample dataset, variables (year and month) are in the rows, which
keeps the data from being tidy. Additionally, the row and column totals
make the task of visualizing and summarizing difficult.

\linespread{1}
\linespread{2}
\vspace{3mm}\setlength{\parindent}{15pt}

The \textbf{reshape} package \citep{R-reshape} was introduced in 2005,
and offered the functions \texttt{melt()} and \texttt{cast()} to perform
data transformations. The package author later realized that the
functions were not consistent with other parts of the tidyverse (such as
\textbf{plyr} \citep{R-plyr}, a predecessor to dplyr), and that the
\texttt{cast()} function needed to be split into two: \texttt{dcast()}
for data frames and \texttt{acast()} for arrays and matrices. These
functions were introduced in the \textbf{reshape2} \citep{R-reshape2}
package in 2010. However, that was not the end of the improvements to
the functions. Many users reported difficulty remembering when to use
\texttt{melt()} and when to use \texttt{cast()}. While they were
somewhat `cute' names, they did not hold inherent meaning in the context
of data analysis. The next iteration in 2014 introduced the functions
\texttt{gather()} and \texttt{spread()} in the tidyr \citep{R-tidyr}
package. The verbs \texttt{gather()} and \texttt{spread()} were somewhat
easier for users to remember, and included the arguments \texttt{key}
and \texttt{value}, which were familiar to database programmers.
However, the improvement left even experienced programmers consulting
the documentation too frequently.

The most recent iteration of the pair of functions is
\texttt{pivot\_wider()} and \texttt{pivot\_longer()}, introduced in
version 1.0.0 of tidyr \citep{R-tidyr}. The names for the functions were
developed as part of a design process that included user surveys. The
resulting functions are much more expressive than the original
\texttt{melt()} and \texttt{cast()} \citep{pivot-tweet}. Instead of
\texttt{key} and \texttt{value}, the arguments use \texttt{names\_to}
and \texttt{values\_to} in \texttt{pivot\_longer()} and
\texttt{names\_from} and \texttt{values\_from} in
\texttt{pivot\_wider()}. Anecdotal evidence (including but limited to
the experience of the authors and many of our students) suggests that
the newest set of functions empower users to write code more fluently
without looking at the full documentation.

Beyond changes to function names, the default values of arguments
present in the tidyverse have been thoughtfully designed with the
intention of making life easier for users and preventing mistakes. These
defaults can be more flexibly updated than those in base R, which, by
design, changes slowly.

While all graphics libraries in R provide for customization, the initial
plots generated by ggplot2 look much more ``finished'' than graphics
from base R or \textbf{lattice} \citep{R-lattice} graphics, which helps
students feel pride in their work from the beginning
\citep{myint2020comparison}. The graphics look more polished because the
defaults have been chosen based on research. For example, the default
ggplot2 color scheme has been updated to use \textbf{viridis}
\citep{R-viridis}, a set of color scales based on perceptual research.
The default grey background is used so that the plot is of similar
visual weight as surrounding text \citep{Wic2021}. When colors or facets
are applied, ggplot2 automatically provides a legend.

The tidyverse provides warnings to analysts to help them avoid making
mistakes. For example, when you create a plot that involves a variable
with missing values, the package will warn you.

\linespread{1}

\begin{Shaded}
\begin{Highlighting}[]
\FunctionTok{ggplot}\NormalTok{(loans) }\SpecialCharTok{+} 
  \FunctionTok{geom\_boxplot}\NormalTok{(}\FunctionTok{aes}\NormalTok{(}\AttributeTok{y =}\NormalTok{ emp\_length, }\AttributeTok{x =}\NormalTok{ application\_type))}
\end{Highlighting}
\end{Shaded}

\begin{verbatim}
## Warning: Removed 794 rows containing non-finite values (stat_boxplot).
\end{verbatim}

\captionof{chunk}{ggplot2 displays a warning when plotting a variable with missing values.}

\label{ggplot-warn} \linespread{2}
\vspace{3mm}\setlength{\parindent}{15pt}

Defaults in other areas of the tidyverse are also designed for success.
The \texttt{drop\_na()} function forces users to think more
intentionally about the way they wish to deal with missing values. In
base R, addressing \texttt{NA} values is either done with a destructive
\texttt{na.omit()} call or as an argument to each function, as shown in
\ref{base-na}.

\linespread{1}

\begin{Shaded}
\begin{Highlighting}[]
\FunctionTok{mean}\NormalTok{(loans}\SpecialCharTok{$}\NormalTok{annual\_income\_joint)}
\end{Highlighting}
\end{Shaded}

\begin{verbatim}
## [1] NA
\end{verbatim}

\begin{Shaded}
\begin{Highlighting}[]
\FunctionTok{mean}\NormalTok{(loans}\SpecialCharTok{$}\NormalTok{annual\_income\_joint, }\AttributeTok{na.rm =} \ConstantTok{TRUE}\NormalTok{)}
\end{Highlighting}
\end{Shaded}

\begin{verbatim}
## [1] 128085.2
\end{verbatim}

\captionof{chunk}{When working with missing data in base R, each function needs an additional argument.}

\label{base-na} \linespread{2} \vspace{3mm}\setlength{\parindent}{15pt}

In the tidyverse, dropping missing values becomes an explicit part of
the pipeline, as shown in \ref{tidy-na}.

\linespread{1}

\begin{Shaded}
\begin{Highlighting}[]
\NormalTok{loans }\SpecialCharTok{|\textgreater{}}
  \FunctionTok{drop\_na}\NormalTok{(annual\_income\_joint) }\SpecialCharTok{|\textgreater{}}
  \FunctionTok{summarize}\NormalTok{(}\FunctionTok{mean}\NormalTok{(annual\_income\_joint))}
\end{Highlighting}
\end{Shaded}

\begin{verbatim}
## # A tibble: 1 x 1
##   `mean(annual_income_joint)`
##                         <dbl>
## 1                     128085.
\end{verbatim}

\captionof{chunk}{Working with missing data in the tidyverse becomes part of the pipeline.}

\label{tidy-na} \linespread{2} \vspace{3mm}\setlength{\parindent}{15pt}

In base R, it is easy to break the relationship between factor levels
and their labels, but the \textbf{forcats} package \citep{R-forcats}
provides many custom functions (prefixed \texttt{fct\_}) for wrangling
categorical data while maintaining sound and reproducible analysis
\citep{mcnamara2018wrangling}. Defaults from the tidyverse---such as not
reading in strings as factor variables---have gained so much popularity
they have been integrated into base R.

We argue that using functionality created to be deliberately
user-centered is vital to bringing new software tools to the classroom.
The more intuitive and memorable the functions, the lighter the
cognitive load for the novice learners we hope to retain
\citep{burr2021, fergusson2021, mcnamaraetal2021a, lovettgreenhouse2000}.
The more functions have been designed with thoughtful defaults, the
easier it is for students to find success.

\hypertarget{sec:reading}{%
\subsection{Readability}\label{sec:reading}}

Programming instruction is improved by reading code out loud
\citep{SwiHer2019}, so it stands to reason that statistics programming
instruction could be similarly improved. People learning to read a human
language (e.g., English) learn by reading aloud and then moving to
``subvocalizing'': saying words in one's head. The reading process
allows learners to connect the sound of the word to the concept it
signifies.

Learning programming by reading aloud can provide similar cognitive
benefits. Students will likely try to subvocalize code as they read
silently, but without examples of phonology (the specific way parts of
the language should be vocalized), they have to make up their own
pronunciations, which may be inconsistent throughout their reading,
adding additional cognitive load \citep{HerSwi2018}. \citet{McN2020}
connects ideas of vocalization to programming using R code,
demonstrating specific phonology for reading R aloud. As with human
language, there can be regional variations in how particular symbols are
voiced, but an instructor should strive to be as consistent as possible
with their choices of vocalizations. Additionally, consistency with
vocalization can help while pair programming or debugging from afar
(e.g., over Zoom), because one person can dictate code to another using
shared language.

The focus on function names as ``verbs'' in the tidyverse lends itself
well to vocalization. Unfortunately, vocalization does not transfer well
to the written page, so the ideas are difficult to convey in the current
manuscript. Consider the sample code in \ref{tidy-vocal} and
\ref{base-vocal}.

\linespread{1}

\begin{Shaded}
\begin{Highlighting}[]
\NormalTok{loans }\SpecialCharTok{|\textgreater{}}
  \FunctionTok{mutate}\NormalTok{(}\AttributeTok{bankruptcy =} \FunctionTok{if\_else}\NormalTok{(public\_record\_bankrupt }\SpecialCharTok{\textgreater{}=} \DecValTok{1}\NormalTok{, }\StringTok{"Yes"}\NormalTok{, }\StringTok{"No"}\NormalTok{)) }\SpecialCharTok{|\textgreater{}}
  \FunctionTok{group\_by}\NormalTok{(bankruptcy) }\SpecialCharTok{|\textgreater{}}
  \FunctionTok{summarize}\NormalTok{(}\AttributeTok{avg\_loan\_amount =} \FunctionTok{mean}\NormalTok{(loan\_amount))}
\end{Highlighting}
\end{Shaded}

\captionof{chunk}{A tidyverse wrangling of the bankruptcy variable in the loans data.}

\label{tidy-vocal} \linespread{2}
\vspace{3mm}\setlength{\parindent}{15pt}

\linespread{1}

\begin{Shaded}
\begin{Highlighting}[]
\NormalTok{loans}\SpecialCharTok{$}\NormalTok{bankruptcy }\OtherTok{\textless{}{-}} \FunctionTok{ifelse}\NormalTok{(loans}\SpecialCharTok{$}\NormalTok{public\_record\_bankrupt }\SpecialCharTok{\textgreater{}=} \DecValTok{1}\NormalTok{, }\StringTok{"Yes"}\NormalTok{, }\StringTok{"No"}\NormalTok{)}
\FunctionTok{tapply}\NormalTok{(loans}\SpecialCharTok{$}\NormalTok{loan\_amount, loans}\SpecialCharTok{$}\NormalTok{bankruptcy, mean)}
\end{Highlighting}
\end{Shaded}

\captionof{chunk}{A base R wrangling of the bankruptcy variable in the loans data.}

\label{base-vocal} \linespread{2}
\vspace{3mm}\setlength{\parindent}{15pt}

Try reading these code snippets out loud to yourself. Which elements do
you vocalize? Which do you skip? When we read the tidyverse code, we
pronounce the \texttt{\textbar{}\textgreater{}} operator as ``and
then.'' ``Start with the loans data and then group by bankruptcy and
then summarize.'' When we read the base code, it is more repetitive
because we need to repeatedly say things like ``loans dollar sign
bankruptcy.'' Typically, we do not vocalize every character on the
screen. Most commonly, we do not read out line breaks, underscores, or
many parentheses (particularly closing parentheses). However, a more
verbose vocalization (perhaps used when dictating to a newer student)
would likely include the additional punctuation. If you would like to
hear one of us read the code snippets aloud, please see
\citet{mcnamara2021}. Hopefully, the exercise of reading code aloud
helps illustrate that the tidyverse is more designed for speakability.
The tidyverse verbs sound more like English sentences (e.g., compare
\texttt{tapply()} to \texttt{group\_by()}).

We should note that readable does not necessarily mean
discoverable---one would not necessarily think of the word ``summarize''
to calculate the mean of a column, but once you learn the framework of
``summarize'' it is likely to stick because the verb does what it says.

Additionally, even when the function names may seem self-explanatory, it
is still important not to assume learners can tell what the function
does without explaining the meaning of the word. \citet{thoma2021}
discusses the importance of the code itself aligning with statistical
frameworks. Tidyverse verbs are based on English, so they privilege
people whose first language is English. And, even for native English
speakers, some of the words are not immediately transferable into the
data analytic context. For example, if you've never come across
\texttt{tidyr::hoist()}, would you be able to guess what ``hoist'' means
in this context? The advantage of well-chosen function names is that
once you explain what a function does, you likely do not need to explain
it again.

\hypertarget{sec:community}{%
\subsection{Community}\label{sec:community}}

We note the tidyverse's popularity across a wide variety of disciplines
and application areas, including in industry. While companies often keep
their software choices private, a number of high-profile companies
publicly use the tidyverse, including Airbnb \citep{bion_how_2018},
T-Mobile \citep{nolis_were_2020}, Stack Overflow
\citep{robinson_exploring_2015}, and many more
\citep{rstudio_pbc_rstudio_nodate}. The tidyverse is also used at many
journalistic outlets, including the BBC
\citep{bbc_visual_and_data_journalism_how_2019} and FiveThirtyEight
\citep{flowers_fivethirtyeights_2016}, as well as by nonprofits such as
the ACLU \citep{watson_r_2019} and The Urban Institute
\citep{dataurban_building_2019}.

Popularity not only indicates that professional users find it to be a
worthwhile tool, but also can actually increase students' ability to
engage with R.
\href{http://www.datasciencemeta.com/rpackages}{DataScienceMeta.com}
tracks downloads of R packages from the Comprehensive R Archive Network
(CRAN) \citep{Rlang}. Six of the top 10 packages (as of July 2, 2021)
downloaded from CRAN are part of the tidyverse, and the tidyverse
package itself is the 20th most downloaded from CRAN. The popularity is
an indication that when students are searching for help (e.g., from
Google or StackOverflow), they are likely to come upon a tidyverse
solution.

Additionally, although older textbooks predominantly use base R to
introduce statistical computing, more and more texts are using tidyverse
syntax. A popular example is \emph{R for Data Science}
\citep{wickham2016r}, a textbook specifically focused on using the
tidyverse to do data science. \citet{ismay2019statistical},
\citet{baumer2021mdsr}, \citet{robacklegler2021},
\citet{hyndmanathanasopoulos2021}, and the R materials associated with
\citet{cetinkaya_hardin_2021} all use primarily tidyverse code.
Providing our students contemporary tools like the tidyverse will
prepare them to engage fully with the larger community of statisticians
and data scientists who have adopted the tidyverse into their work.

\hypertarget{sec:shared-syntax}{%
\subsection{Transferability}\label{sec:shared-syntax}}

The majority of Section \ref{sec:tidyverse} has argued that tidyverse
makes common data analysis tasks more straightforward to learn than
other approaches with R. This reduction in cognitive load will make the
tidyverse easier to learn than base R. Here, we provide one additional
reason for bringing the tidyverse into an undergraduate classroom full
of students who will be heading into a workforce in a data-centered
world.

The careful construction of the tidyverse, and in particular the dplyr
package, can have additional benefits to learners in the context of
working with databases. Since the development of relational database
modeling begun by \citet{codd1970database}, Structured Query Language
(SQL) has been the dominant paradigm for interacting with relational
databases. SQL databases are widely deployed through technologies like
SQLite, MySQL, PostgreSQL, Microsoft SQL Server, and Oracle. Moreover,
many newer technologies that seek to supersede SQL are predicated on
their users' knowledge of SQL. This includes cloud-based services (e.g.,
Google Big Query), as well as non-tabular database systems (e.g.,
``NoSQL''). Thus, for undergraduates, learning how to write SQL queries
is a useful step towards a career in data science
\citep{horton2015taking}, particularly for those headed towards
industry.

The dplyr package was written with SQL in mind. As described in Section
\ref{sec:wrangle}, the main verbs, along with the various
\texttt{*\_join()} functions, comprise a set of functions that can serve
as the building blocks for SQL queries. The dbplyr package provides
functionality that will translate dplyr pipelines into SQL queries,
enabling R users to query SQL databases without having to write SQL
code. However, since knowing how to write an SQL query is a useful skill
for students to develop, learning data wrangling through dplyr has the
beneficial side effect of giving students a conceptual understanding of
SQL with minimal additional cognitive load. That is, the work that
students have put in to learn data wrangling in dplyr can be easily
extended into achieving SQL fluency. Instructors can pair a few weeks of
dplyr instruction with a few weeks of SQL instruction and have
reasonable confidence that students can develop basic proficiency in
\emph{both} technologies.

Students can engage in a comparative literature exercise in which they
map each function in the dplyr pipeline to a different clause in the SQL
statement. The comparison of SQL and dplyr syntax can reinforce the
message that the underlying concepts are the same here: it is only the
programming syntax that differs between R and SQL. To fully drive the
1-to-1 equivalence home, the dbplyr package contains the function
\texttt{show\_query()} that will explicitly translate a dplyr pipeline
to a SQL query. \ref{show-query} shows the SQL translation of the
audiological measurement query shown in \ref{SQL-ex}.

\linespread{1}

\begin{Shaded}
\begin{Highlighting}[]
\NormalTok{db }\SpecialCharTok{|\textgreater{}}
  \FunctionTok{tbl}\NormalTok{(}\StringTok{"Subjects"}\NormalTok{) }\SpecialCharTok{|\textgreater{}} 
  \FunctionTok{group\_by}\NormalTok{(AgeCategoryFirstMeasurement) }\SpecialCharTok{|\textgreater{}}
  \FunctionTok{summarize}\NormalTok{(}\AttributeTok{num\_people =} \FunctionTok{n}\NormalTok{()) }\SpecialCharTok{|\textgreater{}}
  \FunctionTok{show\_query}\NormalTok{()}
\end{Highlighting}
\end{Shaded}

\begin{verbatim}
## <SQL>
## SELECT `AgeCategoryFirstMeasurement`, COUNT(*) AS `num_people`
## FROM `Subjects`
## GROUP BY `AgeCategoryFirstMeasurement`
\end{verbatim}

\captionof{chunk}{A SQL translation of the dplyr pipeline shown in \ref{SQL-ex}.}

\label{show-query} \linespread{2}
\vspace{3mm}\setlength{\parindent}{15pt}

Thus, by learning dplyr students get SQL (almost) for free. We note that
there is also the package \texttt{tidyquery} \citep{R-tidyquery} which
allows for the translation of SQL queries back into dplyr pipelines.

\hypertarget{sec:discussion}{%
\section{Discussion}\label{sec:discussion}}

In this section, we reflect on how the tidyverse fits into a larger
curriculum, discuss the importance and challenge of staying current,
address some common criticisms of our approach, and conclude with final
thoughts.

\hypertarget{building-a-curriculum}{%
\subsection{Building a curriculum}\label{building-a-curriculum}}

We have made the argument that students' first introduction to R can
(and should) be with the tidyverse. However this does not mean learning
materials should be structured around tidyverse packages, as opposed to
statistics and data science concepts. We recommend using
\texttt{library(tidyverse)} to load all eight core tidyverse packages
and not allocating much time or energy to distinguishing which function
lives in which of these eight packages, at least in an introductory
course. It is important to let students know where they can find
information in package documentation, but beyond that, making
distinctions within the tidyverse core packages can add to unnecessary
distractions.

Even just the core eight packages in the tidyverse offer a vast array of
functions for doing data analysis tasks. The breadth of that
functionality goes well beyond the topics that can reasonably be covered
in an introductory statistics or data science course within the span of
an academic term. For example, in introductory courses where the
audience is new to working with data, statistics, and programming, we
recommend delaying introduction of the \textbf{purrr} package
\citep{R-purrr} and the functional programming paradigm. One of the
strengths of the purrr package is working with list-columns, which are
relevant when applying functions to many columns or when working with
hierarchical data. If working with advanced data structures is a topic
included in the learning goals of an introductory course, we recommend
solving the problems using functionality recently added to packages like
dplyr (e.g., the \texttt{across()} function) and tidyr (e.g.,
\texttt{unnest\_*()} functions) in order to avoid introducing functional
programming as an additional topic in the curriculum.

This is not to say incorporating the tidyverse into a curriculum can be
done without any adjustments to the learning goals. For example, if
teaching R without the tidyverse, one might avoid the discussion of the
pipe operator or the notion of a \texttt{tibble} (tidyverse's
implementation of a data frame) entirely. On the other hand, adding some
new learning goals to the course to support the teaching of the
tidyverse can provide a principled framework that allows for tackling
modern data problems while using a consistent syntax.

For an introductory data science course, data visualization is a good
first topic, followed by single-table data wrangling
\citep{cetinkaya2020fresh}. This structure introduces students to
functions from the ggplot2 package and then dplyr, but again we are not
advocating for focusing the course on packages. Subsequent learning
goals for a given course should embrace aspects of tidyverse packages
designed to help with relevant data analysis tasks related to those
goals. For example, multivariate thinking can be introduced alongside
ggplot2 functionality that maps variables to additional aesthetics like
color, size, shape, and facets. Relational data can be introduced with
two-table verbs from dplyr (i.e., \texttt{*\_join()} functions).

\hypertarget{keeping-up-with-the-tidyverse}{%
\subsection{Keeping up with the
tidyverse}\label{keeping-up-with-the-tidyverse}}

Like the majority of (particularly open-source) software, the tidyverse
evolves over time. Many of the changes are responses to feature requests
or difficulties with functions reported by users. The tidyverse team
explicitly solicits feedback from the community on (particularly major)
proposed changes via surveys and blog posts. Changes are generally
announced with each CRAN release of a package in blog posts as well as
in NEWS files of the packages. While the majority of changes are
backwards compatible, a carefully evaluated small subset of them can be
breaking changes \citep{WickhamMaintaining2021}.

The tidyverse uses the \textbf{lifecycle} package to communicate
information on the lifecycle of functions and packages
\citep{R-lifecycle}. Clear messaging via lifecycle badges can help
instructors evaluate whether to teach newly introduced functionality.
For example, one might choose to teach a new function that is in a
\emph{stable} stage but might hold off on an \emph{experimental} one.
Similarly, \emph{deprecated} and \emph{superseded} functions are good
candidates for removal from course materials when revising.

Teaching the tidyverse will therefore always take a little preparation
before class, even if you have materials from a previous term. Because
of occasional breaking changes, you will want to re-run any code you are
providing to students before class. However, quickly reviewing materials
before class is an important practice for any instructor.

\hypertarget{alternative-viewpoints}{%
\subsection{Alternative viewpoints}\label{alternative-viewpoints}}

Some alternative viewpoints to teaching with the tidyverse center on a
general objection to teaching with an excessive number of packages. One
notable description of such minimalism is the
``\href{https://www.tinyverse.org}{tinyverse}.'' There are two main
rationales for these points of view. First, some fear students will
learn a set of programming patterns so specialized, complicated
\citep{leek2016why}, or idiosyncratic \citep{matloff2020tidyverse} that
they will be baffled by the bare R syntax they are likely to see after
the course. (Portions of this argument explicitly support our
characterization of the tidyverse as a coherent syntax.) Second, others
focus on maximizing the durability and robustness of the code written by
minimizing the number of dependencies. Since packages change much more
frequently than R itself, code that relies on packages is more likely to
break in the future \citep{tinyverse2018}. These concerns are logical,
and we address them here.

To the first argument, the tidyverse has become so popular (see Section
\ref{sec:community}) that the fear that tidyverse code will be
unrecognizable, or that students will suffer as a result of their
reliance on the tidyverse, is unfounded. While this objection might hold
for other teaching-focused R packages with small user bases, the
tidyverse ecosystem is too big for this to be a legitimate concern. In
fact, the opposite may be closer to the truth. In recent years we have
seen ideas that were popularized in the tidyverse implemented in base R.
Namely, the change in the default behavior of the
\texttt{stringsAsFactors} argument of \texttt{data.frame()} that
occurred in R 4.0, and the introduction of the native pipe operator
(\texttt{\textbar{}\textgreater{}}) in R 4.1. In our experience,
tidyverse code presented by students to prospective employers is often
seen as evidence of cutting-edge programming skill.

To the second argument, while breaking changes do occur in the
tidyverse, the packages themselves are both well-maintained and
coordinated. \citet{eddelbuettel2018} invokes
\href{https://en.wikipedia.org/wiki/Metcalfe\%27s_law}{Metcalfe's Law}
to argue that as the number of dependencies increases, the probability
that a breaking change in a dependent package will cause a problem also
increases. While this may be true in general, the dependencies of the
tidyverse package are all maintained \emph{by the tidyverse team}, so
the dependencies are not independent---to the contrary, they are highly
correlated. Moreover, while changes in dependent packages can break both
production code and student code alike, the impact of those breakages is
quite different. While robustness is important, there is a complementary
danger of missing out on innovations that will put students in better
positions to succeed. In any case, we view the possibility of these
breaking changes as the price one has to pay for software that is
continually progressing.

Other criticisms of teaching R with the tidyverse to introductory
students center around the tidyverse's extensive use of non-standard
evaluation (NSE). Much of the user-centered design of the tidyverse
relies on the use of NSE within R (e.g., not having to quote column
names within a dplyr function). The complexity of NSE is hidden from
students because it is not something they will meaningfully encounter
until they try to write certain kinds of functions (specifically, a
generic function that uses functions from the tidyverse, which in turn
make use of NSE). This need is unlikely to arise in a first or even
second course in statistics or data science. In advanced courses that
might teach R as a programming language (e.g., package development),
programming with NSE as well as other evaluation patterns used in R can
be covered. Many tidyverse packages provide specific documentation to
help users learn how to use tidyverse tools in functions and packages
they write \citep{programdplyr, ggplot2inpackage}.

\hypertarget{coda}{%
\subsection{Coda}\label{coda}}

We have provided an overview of how the tidyverse works and how it
integrates with undergraduate statistics and data science curricula,
argued that we should start teaching R with the tidyverse, and
articulated core reasons for continuing to use the tidyverse throughout
the curriculum, while touching on features like consistency,
scalability, user-centered design, readability, community, and
opportunities for growth.

We are all converts to the tidyverse and have made a conscious choice to
use it in our research and our teaching. We each learned R without the
tidyverse and have all spent quite a few years teaching without it at a
variety of levels from undergraduate introductory statistics courses to
graduate statistical computing courses. Ultimately, we have settled on
computing curricula that teach (with) the tidyverse and synthesized the
reasons supporting our choice in this paper.

We encourage readers convinced by our arguments to implement the
tidyverse in their classroom teaching. The references that follow
include several textbooks based on the tidyverse, and additional
open-source curricular materials that can be customized or used as-is.
As we have noted, the encouraging and inclusive tidyverse community is
one of the benefits of the paradigm. Welcome! We're glad you're here.

\pagebreak

\bibliographystyle{agsm}
\bibliography{references.bib}

\end{document}